\documentclass[12pt,a4paper]{article}
\usepackage[a4paper,margin=2cm]{geometry}
\usepackage{authblk}

\usepackage{graphicx}  
\usepackage{hyperref}  
\usepackage{mhchem}    
\usepackage{xcolor}    
\usepackage{amsmath}
\usepackage{amssymb}

\newcommand{\beginsupplement}{%
	\setcounter{table}{0}
	\renewcommand{\thetable}{S\arabic{table}}%
	\setcounter{figure}{0}
	\renewcommand{\thefigure}{S\arabic{figure}}%
	\renewcommand{\thesection}{S\arabic{section}}%
}
\title{Graphene Josephson diodes from asymmetric disorder}

\author[1]{Ivan Villani}
\author[2]{Luca Chirolli}
\author[3]{Matteo Carrega}
\author[1]{Alessandro Crippa}
\author[1]{Elia Strambini}
\author[1]{Francesco Giazotto}
\author[4]{Vaidotas Mišeikis}
\author[4]{Camilla Coletti}
\author[1]{Fabio Beltram}
\author[5]{Kenji Watanabe}
\author[6]{Takashi Taniguchi}
\author[1]{Stefan Heun$^{*}$}
\author[1]{Sergio Pezzini$^{*}$}

\affil[1]{Istituto Nanoscienze–CNR, NEST-Scuola Normale Superiore, Piazza San Silvestro 12, 56127 Pisa, Italy}
\affil[2]{Department of Physics and Astronomy, University of Florence, 50019 Sesto Fiorentino, Italy}
\affil[3]{CNR-SPIN, Via Dodecaneso 33, 16146 Genova, Italy}
\affil[4]{Center for Nanotechnology Innovation, Laboratorio NEST, Istituto Italiano di Tecnologia, 56127 Pisa, Italy}
\affil[5]{Research Center for Electronic and Optical Materials, National Institute for Materials Science, 1-1 Namiki, Tsukuba 305-0044, Japan}
\affil[6]{Research Center for Materials Nanoarchitectonics, National Institute for Materials Science, 1-1 Namiki, Tsukuba 305-0044, Japan}

\affil[*]{stefan.heun@nano.cnr.it and sergio.pezzini@nano.cnr.it}

\date{}

\begin{document}
	
	\maketitle

\begin{abstract}
	Josephson diodes are non-reciprocal superconducting devices characterized by different switching currents depending on the current flow direction. They recently attracted considerable theoretical and experimental attention, in view of their possible application as rectifying elements in the field of superconducting electronics, and as probes to investigate symmetry breaking mechanisms in mesoscopic systems. In this work, we show that graphene Josephson junctions provide rectification of supercurrent with an efficiency exceeding $20\%$. The effect appears applying a mT out-of-plane magnetic field and is enhanced close to the nodes of the Fraunhofer interference pattern. Our findings are consistent with a theoretical model that considers long-range scattering potentials in the junction as the symmetry-breaking mechanism, which yields supercurrent rectification in highly transparent junctions. While graphene stands out as an ultra-clean transmission medium, our work shows that unavoidable residual disorder is sufficient to promote this effect.
\end{abstract}

\section{Introduction}

In modern electronics, the standard diode represents a fundamental non-reciprocal device that provides current rectification. The diode finds widespread applications in AC to DC current conversion, circuit protection (Zener diodes), high-speed applications (Schottky diodes), or as light emitting and photo-diodes \cite{pierret2002}. The most common diode implementation is based on a p-n semiconductor junction that explicitly breaks the spatial inversion symmetry.

A supercurrent diode represents the superconducting counterpart of the standard diode: it provides rectification of a supercurrent, that is, for a given range of parameters, it exhibits zero-resistance only for one direction of current flow. The simultaneous breaking of inversion and time-reversal symmetries has been identified as necessary condition for a system to exhibit supercurrent rectification \cite{Ma2025_review, Davydova2022,shaffer2025review}, as this breaks the equivalence between the positive and negative propagation directions.

The supercurrent diode received limited attention until 2020, when Ando and coworkers \cite{Ando2020} observed the effect in a \ce{[Nb/V/Ta]_n} superlattice lacking inversion symmetry. In this case, the origin of the superconducting diode effect lies in the magnetochiral anisotropy induced by the simultaneous breaking of spatial inversion symmetry and time-reversal symmetry. The latter was broken by applying an external in-plane magnetic field perpendicular to the current flow. This anisotropy results in a nonreciprocal critical current, which enables zero-resistance for one direction of current flow.

This first experimental report immediately promoted a tremendous interest in the field, and extensive efforts both on the theoretical \cite{Strambini2020, Zhang2022, Yuan2022, Souto2022,He_2022,Fukaya2025,Nagaosa2024review,Misaki2021,Tanaka2022,Fracassi2024,Cayao2024,fracassi2025,shaffer2025review, Bonasera2026arxiv} and experimental side \cite{Baumgartner2022, Sundaresh2023, Bauriedl2022, Turini2022, Pal2022, Lotfizadeh2024, Costa2023, Baumgartner_2022_arrayJJ, Suri2022, Hou2023, Greco2023, Diez-Merida2023, Margineda2023_1, Li2024, Chen2024, Chen2024_Edelstein, Rashidi2025, Wu2022, Narita2022, Jeon2022, Strambini2022, Lin2022, Chiles2023, Qi2025, Margineda2025_2, Borgongino2025, Zhang2024,Ghosh2024, Satchell2023,Coraiola2024,Gupta2023,Matsuo2023,Valentini2024,Ciaccia2023,Rothstein2026} have been made to investigate and improve a variety of platforms in terms of efficiency and tunability. Furthermore, the diode serves as a powerful tool to identify non-trivial symmetry breaking mechanisms. Several supercurrent diode realizations rely on the application of a magnetic field, both in-plane \cite{Baumgartner2022, Sundaresh2023, Bauriedl2022, Turini2022, Chieppa2025, Lotfizadeh2024, Costa2023, Baumgartner_2022_arrayJJ} and out-of-plane \cite{Suri2022, Hou2023, Greco2023, Diez-Merida2023, Margineda2023_1, Li2024, Chen2024, Chen2024_Edelstein, Rashidi2025}. The effect can also be sustained by synthetic magnetic fields, or obtained in samples with optimized geometric layouts, thus not relying on the application of external fields \cite{Wu2022, Narita2022, Jeon2022, Strambini2022, Lin2022, Gupta2023, Qi2025, Margineda2025_2, Borgongino2025}.

Some of these implementations are based on Josephson junctions, which provide a wide flexibility in terms of geometry, materials choice, and additional tuning knobs such as electrostatic gating and supercurrent modulation. In those cases the diodicity pertains the switching current of the junction, and the phenomenon is known as Josephson diode effect (JDE) -- which differs from the superconducting diode effect (SDE) that affects the critical current of a bulk superconductor as a whole. In Josephson junctions based on materials with large spin-orbit interaction, such as InSb \cite{Turini2022}, InAs \cite{Baumgartner_2022_arrayJJ, Lotfizadeh2024}, or InGaAs \cite{Costa2023}, the JDE emerges from the simultaneous breaking of inversion and time-reversal symmetries through the interplay of Rashba spin-orbit interaction and an external in-plane magnetic field. This combination induces an asymmetric current-phase relation characterized by an anomalous phase shift and cosine harmonics, which ultimately leads to a direction-dependent critical current \cite{Ilic2024}. For the Rashba spin-orbit coupling that breaks mirror symmetry about the direction perpendicular to the junction plane, the JDE takes place under the application of an in-plane magnetic field directed perpendicular to the current \cite{Strambini2020}. The JDE was also reported for gate-defined graphene junctions, when multiple layers are twisted at the so-called magic-angle \cite{Diez-Merida2023, Bhardwaj2025}. This phenomenology is based on correlation-driven phases with valley polarization and orbital magnetism \cite{Hu2023}, that are however obtained at the cost of complex fabrication routes \cite{Lau2022}. Recent results indicate that the JDE in this class of systems can also be driven by disorder, specifically due to twist angle variations along the device \cite{Rothstein2026}.

In this work we focus on Josephson junctions built from a single layer of graphene encapsulated in hexagonal boron nitride (hBN) \cite{Calado_2015}, a platform with negligible spin-orbit coupling \cite{CastroNeto2009}, no correlated electronic phases, and among the cleanest two-dimensional electron systems \cite{Rhodes2019}. We find that these devices exhibit a JDE with rectification efficiency of more than $20\%$ under a purely out-of-plane magnetic field. The two conditions of breaking the time reversal and inversion symmetries, which are necessary (but not sufficient) for the JDE to appear, are satisfied as follows. The time reversal symmetry is explicitly broken by the magnetic field. This also implies that the diode effect is ruled out at $B=0$, which is in agreement with our experimental observations. Regarding inversion symmetry breaking, we consider residual long-range disorder that is present even in graphene/hBN stacks \cite{Decker2011, Rhodes2019}. Guided by a theoretical model originally developed for a two-dimensional electron gas (2DEG)\cite{chirolli2025} and tailored to the case of graphene, we identify a specific mirror symmetry breaking in the scattering potential as necessary condition for the JDE. The relevant vector $\bf{M_x}$ defining this symmetry plane is perpendicular to magnetic field $\bf{B}$ (directed along $z$) and the current direction $\bf{I}$ (so it is parallel to their cross-product: $\bf{M_x}\parallel\bf{I}\times \bf{B}$), as indicated in Fig.~\ref{fig1}a. In a two-dimensional system, this translates to a mirror symmetry breaking about the dashed blue axis in the figure. 

In our experiment, the JDE emerges in the Fraunhofer interference patterns, i.e., in presence of an out-of-plane magnetic field. Fraunhofer patterns differ depending on the direction of the current flow. Supercurrent rectification is observed within the main Fraunhofer lobe and enhanced at the first nodes. Beyond the first nodes, the diode behavior becomes more unpredictable, as it is strongly influenced by the details of the potential landscape.

\section{Experimental results}

\subsection{Investigated devices}

\begin{figure*}[!htb]
	\centering
	\includegraphics[width=\linewidth]{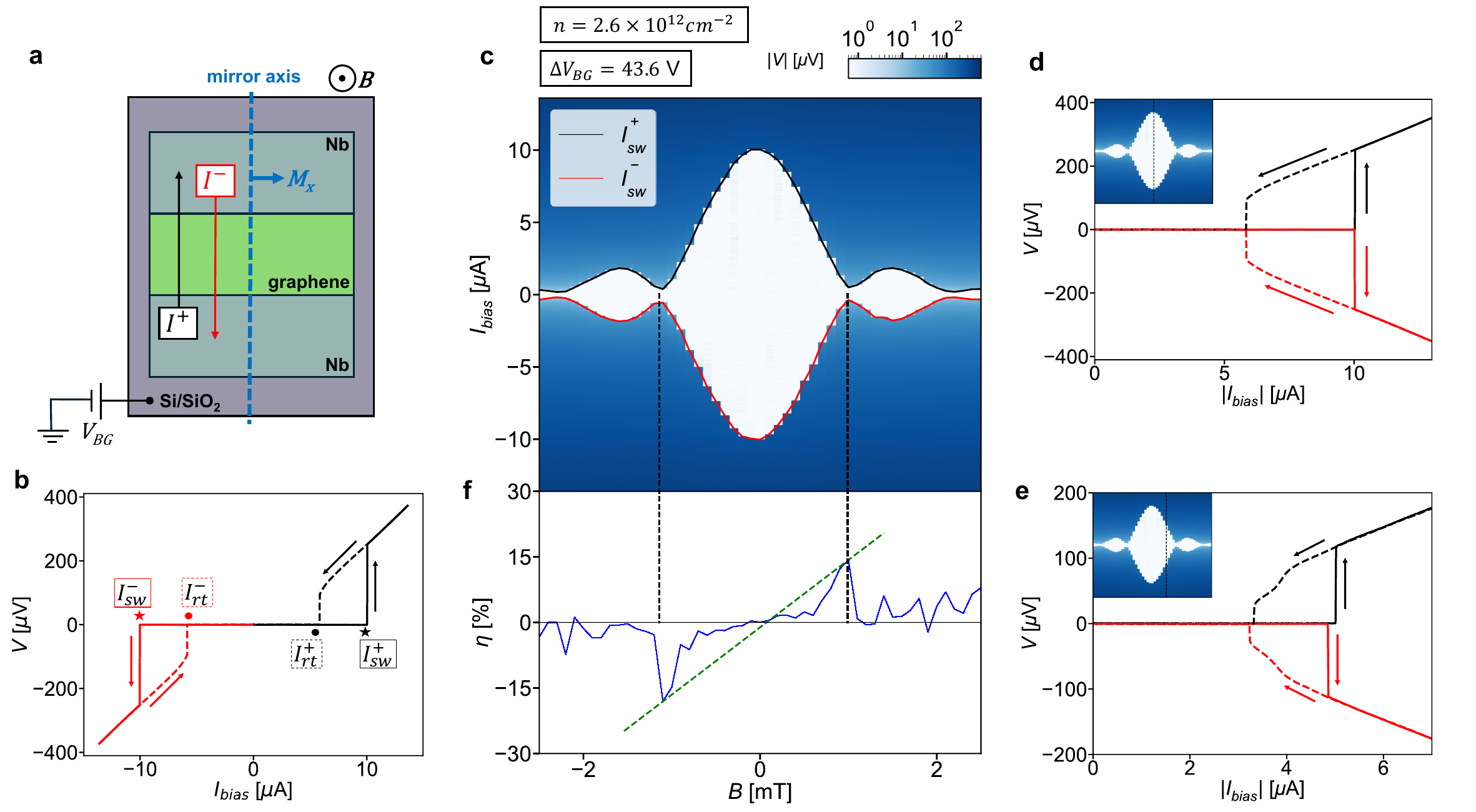}
	\caption{($\mathbf{a}$)
		Top view schematics of device layout. Graphene is encapsulated in hBN and edge-contacted by Nb leads. The heterostructure is placed on top of a \ce{Si/SiO_2} substrate, which acts as a back-gate via the application of a back-gate voltage $V_{BG}$. A current $I$ is applied and flows through the junction, and the voltage drop is measured. Currents flowing in either positive ($I^+$) or negative ($I^-$) directions are indicated in black and red respectively. Magnetic field $B$ is applied in the out-of-plane direction. Mirror symmetry axis $\bf{M_x}$ is indicated. ($\mathbf{b}$) Voltage-current (V-I) curve at $B=0$ showing switching-retrapping behaviour. Positive and negative flowing currents are represented respectively in black and red, with the same notation as in $\mathbf{a}$. The sweep direction of the current is indicated by the arrows. Solid lines indicate the switching branches, while dashed lines indicate the retrapping branches. Positive ($I_{sw}^+$) and negative ($I_{sw}^-$) switching currents are indicated respectively by the black and red stars, while positive ($I_{rt}^+$) and negative ($I_{rt}^-$) retrapping currents by black and red dots. ($\mathbf{c}$) Fraunhofer pattern (voltage drop $|V|$ as function of magnetic field $B$ and DC switching current bias $I_{bias}$): $V_{BG}=40$~V ($\Delta V_{BG}=43.6$~V, $n=2.6\times 10^{12}$~cm$^{-2}$). ($\mathbf{d}$-$\mathbf{e}$) V-I curves taken at $B=0$~mT and $B=0.6$~mT respectively, corresponding to the black dashed lines in the insets, plotted as function of $|I_{bias}|$. Black and red lines indicate respectively positive and negative flowing currents, following the same notation as in $\mathbf{a},\mathbf{b}$, while arrows indicate the sweep direction. In \textbf{d}, no diode effect is observed, i.e., positive and negative switching (and retrapping) currents coincide, while they are different in \textbf{e}. ($\mathbf{f}$) Diode rectification coefficient $\eta$ as a function of magnetic field $B$, calculated using eq.~\eqref{eq:eta}. We define the polarity of the diode as the sign of the slope of the green dashed line, traced by connecting peaks in $\eta$ at first nodes. In this case, the diode has positive polarity. $T=40$~mK for \textbf{b}-\textbf{f}. \label{fig1}}
\end{figure*}

We study the JDE in hBN-encapsulated graphene Josephson junctions contacted by Niobium (Nb) leads and supported by a \ce{Si}/\ce{SiO_2} substrate acting as back-gate (a device schematics is shown in Fig.~\ref{fig1}a). Two Josephson junctions D1 and D2 were studied in this work, fabricated according to the same fabrication protocol (reported in Ref.~\cite{Villani2025}, more details are presented in the  Methods), but belonging to different samples fabricated independently at a time interval of several months. Both devices have length $L=400$~nm and width $W=3\;\mu$m, and are based on high-quality graphene single crystals grown by chemical vapor deposition (CVD) \cite{Miseikis_2015,Miseikis_2017,Pezzini_2020}. Measurements on device D1 were performed in a Leiden Cryogenics dilution refrigerator, at a base temperature of $40$~mK, while measurements on device D2 were performed in a continuous flow ICE \ce{^3 He} refrigerator, at a base temperature of $350$~mK. Both cryostats are equipped with electronic filters anchored to the lowest-temperature plate, comprising double stage RC filters and a $\pi$-filter stage. We report data from device D1 in the main text, while data from device D2 are reported in the Supplementary Material.

The devices show electrical transport characteristics in line with hBN-encapsulated graph\-ene Josephson junctions reported in the literature \cite{Calado_2015, BenShalom2016}. The induced superconductivity is strongly modulated by the voltage applied to the back-gate ($V_{BG}$), as illustrated in the Supplementary Material, Section S1.1. The supercurrent reaches more than 10~$\mu$A thanks to the large interface transparency ($Tr$ up to $0.73$ and $0.84$, for devices D1 and D2, respectively, calculated from the normal state resistance following the procedure outlined in Ref.~\cite{BenShalom2016}; see also Supplementary Material), with comparable values measured for the two devices. Additional characterization of device D1 is provided in the Supplementary Material, Section S1.1. The junctions show a pronounced switching-retrapping behavior, as shown in Fig.~\ref{fig1}b, where we report a representative $V-I$ curve at zero magnetic field, with the corresponding sweep directions indicated by arrows. $I_{sw}^+$ and  $I_{sw}^-$ indicate the switching currents in the positive and negative flow direction, $I_{rt}^+$ and $I_{rt}^-$ the corresponding retrapping currents (the lower value of the retrapping current originates from a self-heating effect\cite{Borzenets2016}). In this work, we use the term \textit{critical current} to identify simulation results, while we use the term \textit{switching current} to identify the experimental results. The measured switching current is used as an experimental estimate of the \textit{critical current} of the Josephson junction.

\subsection{JDE observation}
A Fraunhofer interference pattern is observed upon application of an out-of-plane magnetic field, as expected for extended junctions, resulting from the interference of the supercurrent as magnetic flux threads the junction area. In our devices, Fraunhofer patterns are different depending on the direction of current flow. This is a manifestation of the JDE. The effect is observed within the main and first lateral lobes. Therefore, all Fraunhofer acquisitions reported here were limited to a $\pm2.5$~mT range (Fraunhofer patterns of device D1 in an extended $\pm10$~mT range are shown in the Supplementary Material, Section S1.1). In Fig.~\ref{fig1}c we show the Fraunhofer interference pattern at large electron doping, {$\Delta V_{BG}=43.6$ V}. $\Delta V_{BG}$ is defined relative to the charge neutrality point (CNP), $\Delta V_{BG}=V_{BG}-V_{BG}^{CNP}$ ({$V_{BG}=40$ V} in Fig.~\ref{fig1}c). For device D1, $V_{BG}^{CNP}=-3.6$~V, a typical value for Nb-contacted graphene \cite{BenShalom2016}. The data in the color map is obtained by sweeping the current bias from zero, highlighting the positive and negative switching currents (black and red lines, respectively). The JDE is illustrated in detail in Figs.~\ref{fig1}d,e, where we report the positive and negative switching and retrapping sweeps from the pattern shown in Fig.~\ref{fig1}c, at two different magnetic field values. In Fig.~\ref{fig1}d, taken at $B=0$ (black dashed line in the inset of Fig.~\ref{fig1}d), $I_{sw}^+=|I_{sw}^-|$ and $I_{rt}^+=|I_{rt}^-|$, i.e., no JDE is observed. Conversely, in Fig.~\ref{fig1}e, taken at $B=0.6$~mT (black dashed line in the inset of Fig.~\ref{fig1}e), $I_{sw}^+\neq |I_{sw}^-|$ and $I_{rt}^+ \neq |I_{rt}^-|$, hence we observe supercurrent rectification. We remark that the JDE is observed in both switching ($I_{sw}$, continuous lines in Fig.~\ref{fig1}b) and retrapping directions ($I_{rt}$, dashed lines in Fig.~\ref{fig1}b). The lower value of the retrapping current is caused by a self-heating effect (thus, the mechanism for supercurrent rectification is the same for switching and retrapping currents). In the following discussion on JDE, we focus on the switching current. Comparable results are obtained for the retrapping current and are shown in the Supplementary Material, Section S1.2. 

The efficiency of a supercurrent diode is given by the rectification parameter $\eta$, defined as the ratio between the difference and the sum of the switching currents propagating in opposite directions:
\begin{equation}
	\eta=\frac{I_{sw}^+-|I_{sw}^-|}{I_{sw}^++|I_{sw}^-|}
	\label{eq:eta}
\end{equation}

The diode efficiency $\eta$ for the data shown in Fig.~\ref{fig1}c is reported in Fig.~\ref{fig1}f as a function of magnetic field $B$. The rectification parameter shows a characteristic anti-symmetric shape, indicating that time-reversal symmetry is broken only for $B\neq0$, and that a reversal of the magnetic field maps $I_{sw}^+$ onto $I_{sw}^-$ and vice-versa. $\eta$ is enhanced at the field values corresponding to the first Fraunhofer nodes, and is, in this case, negligible within the first lateral lobes. We note that even at the nodes $I_{sw}^+$ and $I_{sw}^-$ remain of the order of several hundreds of nA. In this context, a relevant complementary metric to assess the diode performance is provided by the absolute switching current difference $\Delta I_{sw}=I_{sw}^+-|I_{sw}^-|$. For the case presented in Fig.~\ref{fig1}c, at the Fraunhofer node at $B=-1.1$~mT, $I_{sw}^+=385$~nA, $I_{sw}^-=-555$~nA, resulting in $\Delta I_{sw}=-170$~nA. The maximum $\Delta I_{sw}=-453$~nA, instead, is observed at slightly lower magnetic field ($B=-0.8$~mT), where the switching currents already exceed $3\;\mu$A. Additional data and discussion are provided in Section S1.3 of the Supplementary Material. We define the polarity of the diode as the sign of the slope of the line obtained by connecting the peaks in $\eta$ at the first nodes. For the case presented in Fig.~\ref{fig1}f, the sign of the slope of the green dashed line is positive, so the diode has positive polarity.

\begin{figure*}[!htb]
	\centering
	\includegraphics[width=\linewidth]{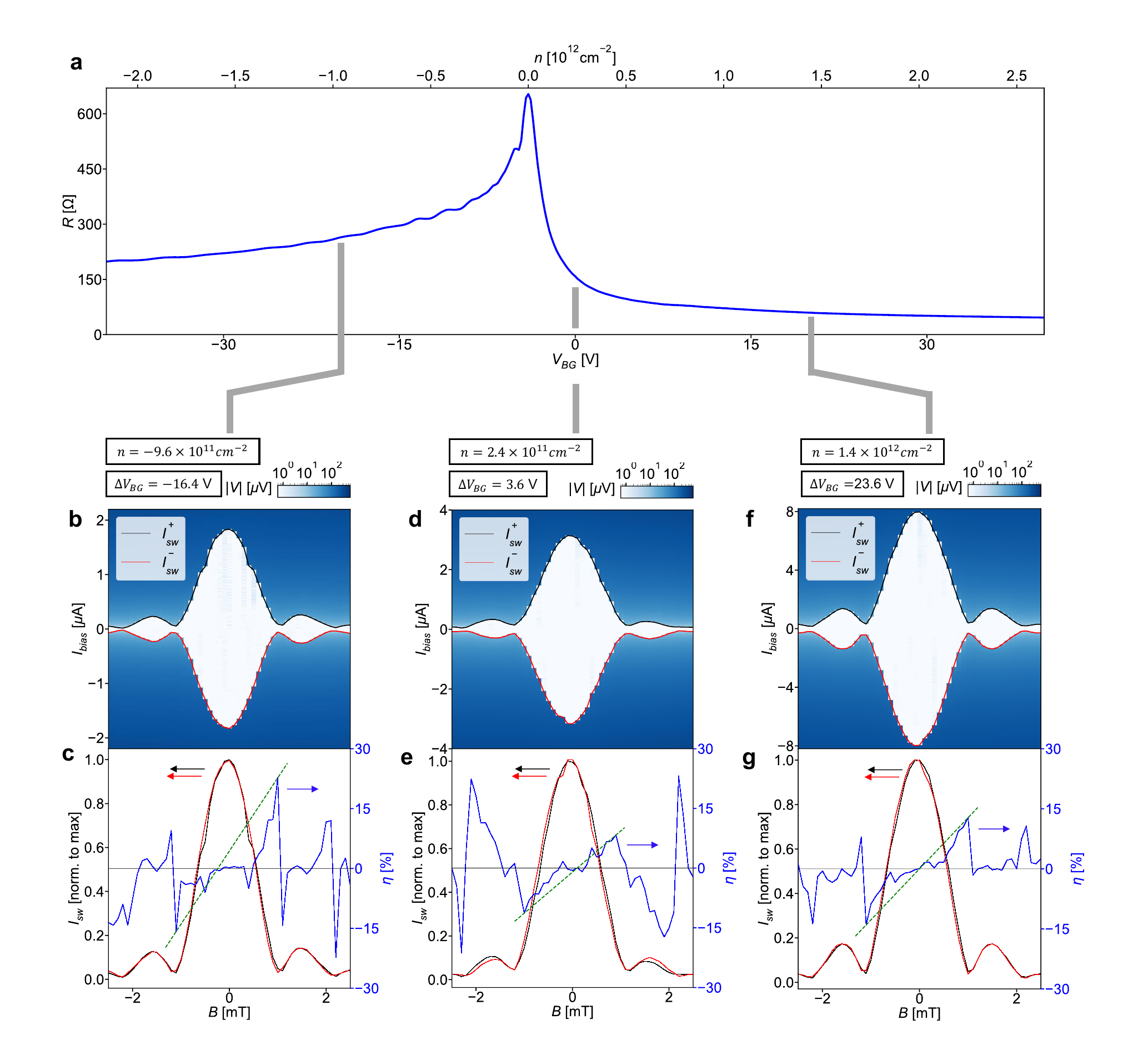}
	\caption{(\textbf{a}) Back-gate sweep: sample resistance $R$ as a function of back-gate voltage $V_{BG}$ (lower x-axis), measured by applying a large DC current bias. The top x-axis shows the corresponding charge carrier density $n$. The acquisition position of the Fraunhofer patterns in \textbf{b},\textbf{d},\textbf{f} is indicated. $T=4.2$~K, $B=0$~T. (\textbf{b}-\textbf{g}) \textit{Upper row}: Fraunhofer patterns (voltage drop $|V|$ as function of out-of-plane magnetic field $B$ and DC bias current $I_{bias}$). \textit{Lower row}: corresponding rectification parameter $\eta$ at different back-gate voltage values. The relative position to the charge neutrality point {$\Delta V_{BG}=V_{BG}-V_{CNP}$} ($V_{CNP}=-3.6$~V) and the corresponding charge carrier density values are indicated at the top of each panel. (\textbf{b}-\textbf{c}) $V_{BG}=-20$~V. (\textbf{d}-\textbf{e}) $V_{BG}=0$~V. (\textbf{f}-\textbf{g})  $V_{BG}=+20$~V.  \label{fig2}}
\end{figure*}

In Figs.~\ref{fig2}b,d,f we report Fraunhofer patterns at different back-gate voltages, respectively at $V_{BG}=-20,0,+20$~V, at the positions indicated in the back-gate sweep of the normal-state resistance $R$ shown in Fig.~\ref{fig2}a. Due to the aforementioned $V_{BG}^{CNP}=-3.6$~V, the junction is hole-doped at $V_{BG}=-20$~V, while for $V_{BG}=0,+20$~V, the junction is electron-doped. This set of measurements thus allows to evaluate the JDE as the carrier density is varied. The carrier density $n=C_g(V_{BG}-V_{BG}^{CNP})$, where $C_g=6\times10^{10}\text{cm}^{-2}\text{V}^{-1}$ is the back-gate lever arm for our gating configuration \cite{Pezzini_2020}, which is also reported as the top x-axis in Fig.~\ref{fig2}a and for each Fraunhofer pattern. In Fig.~\ref{fig2}c,e,g we report the corresponding rectification coefficients (blue lines), along with positive and negative switching currents (black and red lines) normalized to the maximum value (since the supercurrent is modulated by $V_{BG}$, the maximum supercurrent is different for the three patterns). All three patterns exhibit supercurrent rectification within the main Fraunhofer lobe, with enhanced $\eta$ at the nodes, and they share the same diode polarity. However, differences are present, with larger maxima in the diode efficiency at the first node visible in Fig.~\ref{fig2}b,c, where $\eta$ exceeds $20\%$. In Fig.~\ref{fig2}d,e, at $V_{BG}=0$~V, we obtain the JDE notably in gate-free operation. In this case, $\eta$ is clearly non-zero also on the first lateral lobes, while negligible rectification is observed on the first lateral lobes in Fig.~\ref{fig2}f,g (similar to Fig.~\ref{fig1}c,f). As we will show in the following, this variability is captured by our theoretical model and determined by the interplay of microscopic details in the disorder and the chemical potential set by $V_{BG}$. Similar results are obtained for device D2 and for an additional device D3 with different geometry, presented in the Supplementary Material, Sections S2 and S3.

\section{Theoretical model}

The theoretical description of the JDE in the Fraunhofer patterns of planar Josephson junctions has been provided in Ref.~\cite{chirolli2025} for the case of a 2DEG. It was shown that the diode effect is originated by the breaking of the mirror symmetry of the junction about the $\bf{M_x}$ axis due to an asymmetry in the scattering potential in the central region of the junction. This is sufficient to yield the diode effect in highly transparent Josephson junctions, in combination with an out-of-plane magnetic field which breaks the time-reversal symmetry. All relevant details about the scattering matrix model are summarized in the Supplementary Material, Section S4. 

For the present case, the theoretical model has been adapted to the case of graphene. We model the graphene sheet by a tight-binding Hamiltonian:
\begin{equation}\label{Eq:Hamiltonian_graphene}
	H=-t\sum_{<ij>s}(e^{i\gamma_{ij}}c^\dag_{is}c_{js}+{\rm H.c.})+\sum_{js}(V_{j}-\mu)c^\dag_{js}c_{js},
\end{equation}
where $c_{js}$ is a fermionic annihilation operator describing an electron with spin $s$ at site $j$, which hops in a honeycomb lattice and picks up a phase $\gamma_{ij}=\frac{e}{\hbar}\int_{{\bf r}_i}^{{\bf r}_j}d{\bf s}\cdot{\bf A}({\bf s})$ due to a uniform magnetic field ${\bf B}$ described by the vector potential ${\bf A}=\frac{1}{2}{\bf r}\times {\bf B}$ in the symmetric gauge. $\mu$ is the chemical potential and $V_j\equiv V(x_j,y_j)$ is a potential experienced by electrons and describing the effect of long range disorder. The back-gate is assumed to globally change the chemical potential $\mu$. In pristine graphene, $t=2.7$ eV is the nearest neighbor hopping energy \cite{CastroNeto2009}. In the simulations, $t$ is a reference energy scale. Given the size of the simulated system with an effective lattice spacing $a$, the low-energy dynamics around the Dirac point is recovered by setting the Dirac velocity to $v_D=3at/2$.

\subsection{Highly doped junctions with different disorder}

\begin{figure*}[!htb]
	\centering
	\includegraphics[width=\linewidth]{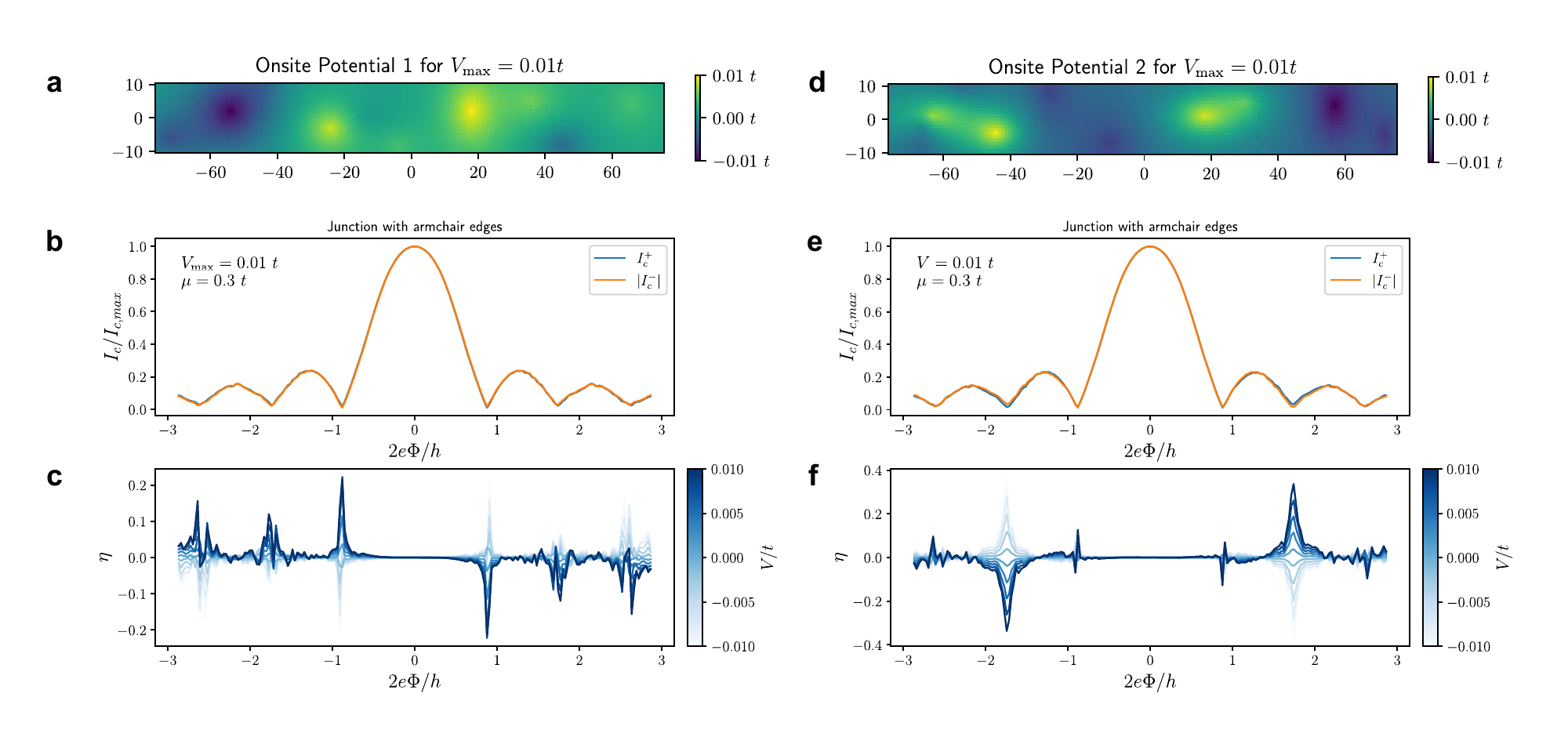}
	\caption{($\mathbf{a}$) Onsite potential $V_1({\bf r})=Vf_1({\bf r})$ for $V=V_{\rm max}=0.01~t$. The spatial coordinates are in units of the effective lattice spacing. (\textbf{b}) Calculated Fraunhofer patterns ($I_c^+$ and $|I_c^-|$) for $V=V_{max}$ and chemical potential $\mu=0.3 t$. (\textbf{c}) Color-coded evolution of the rectification coefficient $\eta$ for varying $V$ in the range $-V_{\rm max}<V<V_{\rm max}$, as a function of applied magnetic flux $2e\Phi/h$. (\textbf{d}) Onsite potential $V_2({\bf r})=Vf_2({\bf r})$ for $V=V_{\rm max}=0.01~t$. (\textbf{e}-\textbf{f}) Same as \textbf{b},\textbf{c}, but for the onsite potential $V_2$ shown in \textbf{d}. \label{fig3_simul_var_dis}}
\end{figure*}

Far away from the Dirac point, a doped graphene sheet behaves very much like an ordinary 2DEG, with the only difference that the sites $j$ are now on a honeycomb lattice and the Hamiltonian is that of Eq.~\ref{Eq:Hamiltonian_graphene}. In Fig.~\ref{fig3_simul_var_dis}a,d we report two different potentials, $V_1({\bf r})=Vf_1({\bf r})$ and $V_2({\bf r})=Vf_2({\bf r})$, shown for $V=V_{max}=0.01t=27$~meV as a function of spatial coordinates, expressed in units of the effective lattice spacing. Here, $f_1(\bf{r})$ and $f_2(\bf{r})$ are two different non-$\bf{M_x}$ symmetric functions describing the spatial evolution of the potential landscape. They are chosen such that $\max_{\bf r}f_i=-\min_{\bf r}f_i=1$, and the geometrical aspect ratio $x/y$ is the same as for the measured devices ($W/L=7.5$). $V_{max}$ is set to match the energy scale of the typical charge-fluctuation density of encapsulated graphene devices. For device D1 we obtain a charge-fluctuation density $n^*=3.4\times 10^{10}$ cm$^{-2}$ that corresponds to $22$~meV according to the Dirac relation \cite{CastroNeto2009} (for D2 $n^*=1.9\times 10^{10}$ cm$^{-2}$, corresponding to $16$~meV); $n^\star$ values are obtained using the method outlined in Ref.~\cite{Pezzini_2020} and are in line with previous reports of high-quality graphene Josephson junctions \cite{Rhodes2019,BenShalom2016}. It is therefore reasonable to expect the disorder potential in our devices to be analogous to other state-of-the-art hBN-encapsulated graphene Josephson junctions \cite{Calado_2015,BenShalom2016,Borzenets2016}. Further details are provided in the Supplementary Material. The two simulated scattering potentials have a different long-range scattering center layout, but comparable characteristic spatial scales, with maxima and minima alternating on the scale of tens of effective lattice constants, corresponding to few hundreds of nm. In the resulting Fraunhofer patterns shown in Figs.~\ref{fig3_simul_var_dis}b,e, different $I_{c}^+$ and $|I_{c}^-|$ are visible.

In Fig.~\ref{fig3_simul_var_dis}c,f we show the rectification coefficient $\eta$ calculated for different values of the amplitude $V$ of the potential landscape $-V_{\rm max}<V<V_{\rm max}$. Results in Fig.~\ref{fig3_simul_var_dis}c are obtained for the potential $V_1$, while in Fig.~\ref{fig3_simul_var_dis}f we show the rectification induced by the potential $V_2$. For the two potentials, the rectification parameter evolves in different ways with magnetic flux, showing that a device-specific pattern is to be expected related to the microscopic configuration of the scattering potential. Moreover, the effect always retains certain general features. For both potentials, a non-zero rectification coefficient $\eta$ is obtained, which is enhanced in proximity to the nodes of the Fraunhofer pattern and antisymmetric around B=0. Our experimental results show a rectification coefficient $\eta$ that is either finite (Fig.~\ref{fig2}e) or very low (Figs.~\ref{fig1}d,\ref{fig2}c,g) within the main Fraunhofer lobe, and antisymmetric around $B=0$. In all cases, it is enhanced at the first Fraunhofer nodes, consistent with the theoretical simulations shown in Fig.~\ref{fig3_simul_var_dis}.

Although we cannot directly tune the scattering potential in the experiment, we can compare the JDE in different devices with different microscopic realizations of the disorder. Since the normal-state characteristics $R(V_{BG})$ and the charge fluctuations $n^*$ are comparable for devices D1 and D2 (see Fig.~\ref{fig2}a and Supplementary Material Figs.~S1e,~S5a,g), the amplitude of the scattering potential is expected to be similar as well. A notable difference lies however in the shape of the Fraunhofer patterns of the two devices. While for D1 they appear to be more irregular (see Supplementary Material, Section S1.1), for D2 they are closer to the expected $\sin(x)/x$ form (see Supplementary Material, Section S2.1). It is therefore reasonable to assume that the two devices provide two different microscopic realizations of the disorder potential. Despite this difference, they exhibit a similar JDE -- both in terms of magnitude of the rectification coefficient $\eta$ and its dependence on the magnetic field -- in accordance with our simulations considering scattering potentials with different shapes and amplitude.

\subsection{Dependence on edge termination and chemical potential}

\begin{figure*}[!htb]
	\centering
	\includegraphics[width=\linewidth]{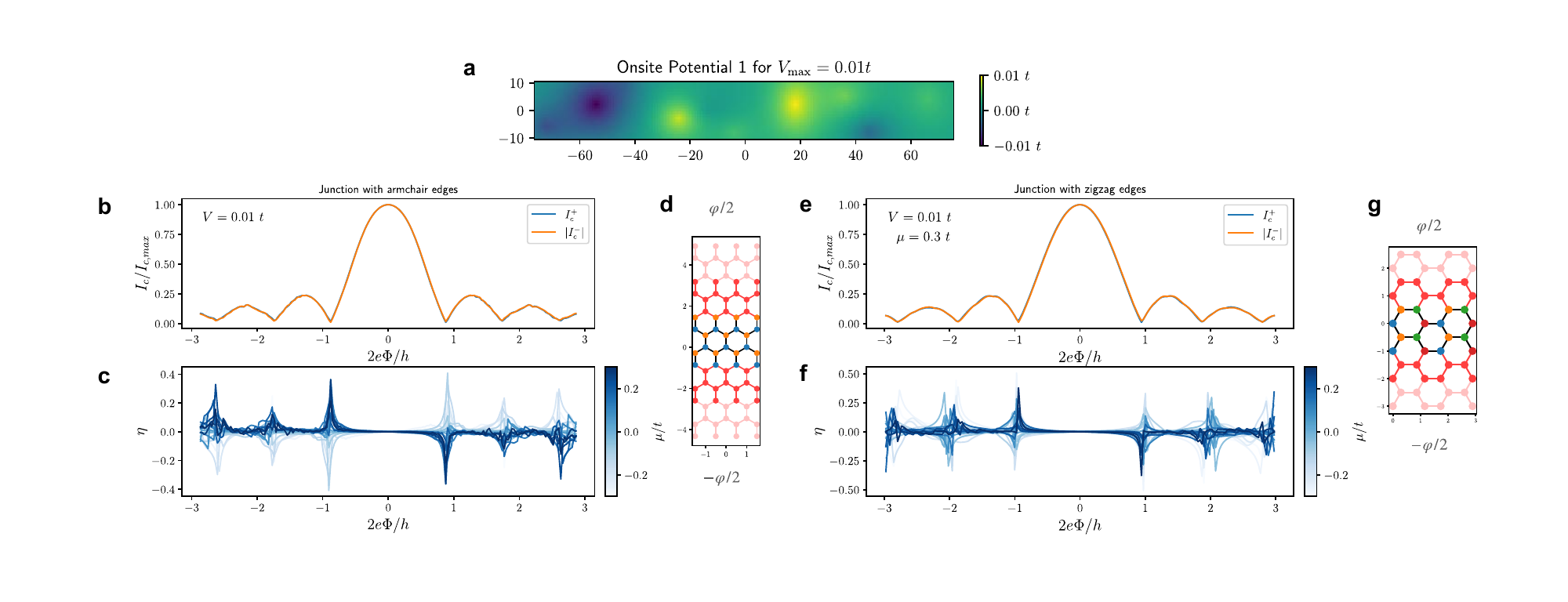}
	\caption{($\mathbf{a}$) Onsite potential $V_1$ for $V=V_{\rm max}=0.01~t$. (\textbf{b}) Calculated Fraunhofer patterns ($I_c^+$ and $|I_c^-|$) for the case of armchair edges for $\mu=0.3 t$. (\textbf{c}) Color-coded evolution of the rectification coefficient $\eta$ with chemical potential $\mu$, crossing the Dirac point. (\textbf{d}) Graphene strip with armchair edges. (\textbf{e}-\textbf{f}) Same as \textbf{b}-\textbf{c}, but for a graphene strip with zigzag edges as shown in \textbf{g}. \label{fig4_var_mu}}
\end{figure*}

In Fig.~\ref{fig4_var_mu} we plot the simulated Fraunhofer patterns for the same potential $V_1$ as in Fig.~\ref{fig3_simul_var_dis}a, shown in Fig.~\ref{fig4_var_mu}a, for two nominally identical graphene sheets, only differing by their relative orientation, with armchair edges in (b,c), as schematized in (d), and with zigzag edges in (e,f), as schematized in (g). In Fig.~\ref{fig4_var_mu}b,e the chemical potential is set to $\mu=0.3~t$ and Fraunhofer patterns with different $I_{c}^+$ and $|I_{c}^-|$ are obtained. In Fig.~\ref{fig4_var_mu}c,f we study the variation of the rectification coefficient $\eta$ with chemical potential $\mu$ varying across the Dirac point, located at $\mu=0$. Comparing Figs.~\ref{fig4_var_mu}c,f, we note that no relevant differences are observed that would depend on the relative orientation of the graphene edges. We remark that, by using common plasma etching techniques, this parameter cannot be experimentally controlled, resulting in edges comprising a random combination of the two terminations.

The rectification coefficient in Figs.~\ref{fig4_var_mu}c,f shows enhanced values around all nodes of the Fraunhofer pattern and variations with the chemical potential. We experimentally observe similar variations in our devices as $V_{BG}$ is changed. For example, $\eta$ is clearly non-zero in the first lateral lobes in Fig.~\ref{fig2}d, while its value is very small in other cases. It is important to note that the character of the variation with $\mu$ does not always show effective tunability with the sign of the chemical potential. This is reasonable, in that a change of sign of the chemical potential maps electron-like excitations to hole-like excitations, which experience a potential $-V$ of opposite sign. The latter is in general different from its mirror symmetric form $M_x^{-1}VM_x$ ($M_x$ is the operator performing a mirror symmetry operation about the $\bf{M_x}$ plane in Fig.~\ref{fig1}a), and a change of sign of the rectification coefficient is therefore not expected. Indeed, although in Fig.~\ref{fig4_var_mu}c we see a change of sign of the rectification with $\mu$, this is not evident in Fig.~\ref{fig4_var_mu}f. This is in agreement with our experimental observations, where no change in the diode polarity is observed in device D1 while crossing the CNP (see Figs.~\ref{fig2}c,e), while it is observed in device D2, as shown in the Supplementary Material, Section 2.2. The polarity of these graphene Josephson diodes is determined by the details of the scattering potential, which cannot be directly controlled by acting on the back-gate. As discussed in the Supplementary Material, Section S5, a deterministic diode polarity flip is obtained by performing a mirror symmetry operation on the scattering potential about the $\bf{M_x}$-axis, but this control is experimentally challenging.

We next consider the case of chemical potential $\mu$ close to the Dirac point. This regime is dominated by strong density inhomogeneities, known as charge puddles \cite{Samaddar2016}. Their formation drastically modifies the transport properties as compared to highly doped graphene \cite{rmp}. A Fraunhofer pattern measured close to the CNP is shown in Supplementary Material Fig.~S4. We observe a distortion of the interference pattern, with a partial merging of the main and first side lobes (i.e., a suppression of the first minima). Such shape can be qualitatively connected to changes in the supercurrent distribution, which could be reasonably expected in the charge puddle regime. However, when considering the JDE, there is no reason to expect substantial differences in the rectification coefficient as compared to the highly doped case. The diode is caused by an asymmetry in the electrostatic potential, which is present even in low doping regimes. Both in the experimental data and in the simulations, at low doping $\eta$ exhibits comparable behavior to the large doping case in terms of amplitude and magnetic field dependence.

\section{Discussion}

To observe supercurrent rectification, it is necessary to simultaneously break the inversion and time-reversal symmetries. A common way to achieve this is by exploiting Rashba spin-orbit interaction in combination with an in-plane magnetic field. Here we can rule out this mechanism because monolayer graphene presents negligible spin-orbit coupling \cite{Yugui2007}. We also do not expect any relevant contribution arising from interaction with the hexagonal Boron Nitride substrate \cite{Island2019, Rout2024}. We can also exclude effects such as valley polarization and orbital magnetism found in moiré multilayers \cite{Diez-Merida2023,Bhardwaj2025,Hu2023}. Moreover, we obtain supercurrent rectification without any specific geometrical engineering (such as the trapezoidal devices analyzed in Ref.~\cite{chirolli2025} or the Josephson triodes presented in Refs.~\cite{Wu2022,Chiles2023}) that would lead to a geometrically induced supercurrent rectification. Finally, we recall that previous reports \cite{Chen2024,Rashidi2025} of JDE in out-of-plane magnetic fields identify the self-field generated by the supercurrent as the symmetry breaking mechanism. In this case, the JDE is predicted to be enhanced as the switching current increases, but this is not consistent with our experimental observations. In fact, we observe maximum rectification at the minima of the Fraunhofer patterns. Here the self-field induced by the supercurrent is minimum and thus much smaller than close to the Fraunhofer maximum, where instead suppressed supercurrent rectification is observed. This is in contrast with the prediction of a larger rectification due to a larger self-field, and we therefore discard this explanation.

\section{Conclusions}
We have demonstrated that hBN-encapsulated graphene Josephson junctions exhibit supercurrent rectification upon the application of an out-of-plane magnetic field. Supercurrent rectification is observed in the Fraunhofer patterns, irrespective of doping level and carrier sign. All our experimental observations are consistent with an asymmetric scattering potential as the source for breaking of the mirror symmetry, enabling the JDE. Simulations of the rectification coefficient $\eta$ under different conditions match well the experimental observations. In our simulations, we explored different potential shapes that break the required symmetries, all qualitatively similar but quantitatively different. Supercurrent rectification is observed in the main Fraunhofer lobe and enhanced at the first nodes. However, the details of the rectification coefficient depend on the shape of the potential, on its amplitude, and on the chemical potential. Such variability does not follow a specific trend, but it rather evolves in an uncontrolled way, which is consistent with the experimental data. Indeed, different devices set to different back-gate voltages realize a variety of configurations in terms of potential shape, amplitude and chemical potential: for this reason, quantitative agreement between experimental data and the simulations is not to be expected. However, the general behavior of $\eta$ is consistently preserved: $\eta$ is always antisymmetric around $B=0$ and enhanced at the nodes, where it reaches values of approximately $20\%$. Beyond the first nodes, the character of the rectification coefficient depends crucially on the details of the scattering potential, as expected from the theoretical simulations. The diode polarity does not appear to be determined by the sign of the charge carriers, consistent with a non-polar nature of the scattering potential but is only affected by the details of the disorder, and for this reason, in our devices it is not directly controllable.

Our devices provide an easy-to-access proof-of-concept platform for supercurrent rectification, only requiring an out-of-plane magnetic field as external knob to exhibit the effect, and potentially serving as a probe to study symmetry breaking mechanisms in this mesoscopic system. Further engineering may provide the necessary elements for additional tunability for applications in the realm of superconducting electronics. For example, direct control of the scattering potential could be implemented by the use of additional side gates. These have been demonstrated on InSb nanoflag Josephson junctions \cite{Guiducci2019,Turini2025} and on graphene-based Josephson junctions \cite{Seredinski2019} to be highly effective in tuning transport properties, not only in the normal regime, but also in the dissipation-less regime. They could be used to deterministically change the shape and symmetry of the scattering potential, thus possibly serving as tuning knobs for the JDE and its polarity. In conclusion, graphene Josephson junctions in mT magnetic fields represent a valid JDE platform, and we expect they could be further engineered to achieve broader and more reliable tunability of the JDE.

\section{Methods}
Fabrication of graphene Josephson junction devices starts from the assembly of the hBN-graphene-hBN stack, performed by employing the standard dry pick-up technique described in Refs.~\cite{Wang2013, Purdie2018,Pezzini_2020}. A poly(bisphenol A carbonate) (PC) film is deposited onto a polydimethylsiloxane (PDMS) \cite{Purdie2018} stamp, which is attached to a glass slide. This stamp is used to combine micromechanically exfoliated hBN flakes ($\sim30$ nm thick) and CVD-grown monolayer graphene single-crystals \cite{Miseikis_2015, Miseikis_2017}, as described in Ref.~\cite{Pezzini_2020}. After the assembly, Raman spectroscopy is used to identify target areas characterized by minimal nanoscale strain variations \cite{Couto2014, Neumann2015, Pezzini_2020}, in which devices are then located. Device patterning is done through electron beam lithography (EBL) employing a polymethylmethacrylate (PMMA) mask (AR-P 672.045 resist). The first EBL step defines a mask which is used for both etching and sputtering, leading to self-aligned one-dimensional superconducting contacts \cite{BenShalom2016}. First, a $15$~s mild oxygen plasma step ($10$ W power) is performed prior to etching to remove  polymer residues in the exposed areas. A mixture of \ce{CF_4} and \ce{O_2} (with flowrates of 20 sccm and 2 sccm, respectively, resulting in a process pressure of $\sim 6\times 10^{-2}$ mbar) is then used to etch through the entire $60$ nm-thick stack in approximately $30$ s at $25$ W power. An additional $10$ s oxygen plasma step, which follows the \ce{CF_4}-\ce{O_2} etching, yields an increased Nb-graphene interface transparency (by typically $20-30$~\%). $60$ nm-thick Nb contacts are then deposited using DC magnetron sputtering at a rate of $\sim 1$~nm/s, with no adhesion layer deposited prior to Nb. The thickness of the Nb is chosen to match the hBN-graphene-hBN stack thickness. Lift-off is performed in acetone at a temperature of $T=50^\circ$C for 20 minutes. A second EBL step defines the device mesa: the same RIE process used for the contacts is employed to etch the area around the devices and to define their geometry. Sample fabrication is completed with an overnight cleaning in acetone at room temperature, to remove  polymer residues.

\section*{Supplementary Material}
Additional characterization of device D1: BG-modulation of the supercurrent, Multiple Andreev Reflections, calculation of interface transparency, charge fluctuations (Fig. S1), extended Fraunhofer patterns of device D1 at different BG voltages (Fig. S2), JDE on the retrapping current, additional metric $\Delta I_{sw}$ for the JDE (Fig. S3), rectification efficiency as a function of different threshold values on $I_{sw}$ (Table S1), diode effect at the CNP (Fig. S4); data on device D2: device characterization (Fig. S5), diode effect (Fig. S6); supercurrent rectification on additional device D3 (Fig. S7); additional theory: details on scattering matrix formulation, diode polarity flip as a result of a mirror symmetry operation on the scattering potential (Fig. S8).

\section*{Acknowledgments}
We acknowledge financial support from the PNRR MUR Project PE0000023-NQSTI funded by the European Union-NextGenerationEU. E.S. acknowledges the Italian Project HELICS DFM.AD002.206. F.G. acknowledges the EU’s Horizon 2020 Research and Innovation Framework Programme under Grants No. 964398 (SUPERGATE) and No. 101057977 (SPECTRUM) for partial financial support. K.W. and T.T. acknowledge support from the CREST (JPMJCR24A5), JST and World Premier International Research Center Initiative (WPI), MEXT, Japan.

\section*{Data availability}
	
The data that support the findings of this study are openly available at the following URL/DOI: \url{https://doi.org/10.5281/zenodo.22128961}

\clearpage
\appendix
\beginsupplement
\setcounter{section}{0}
\setcounter{figure}{0}
\setcounter{table}{0}
\setcounter{equation}{0}

\begin{center}
	\section*{Supplementary Material}
\end{center}

\section{Additional data from device D1}

\subsection{Additional characterization}

\begin{figure*}[!htb]
	\centering
	\includegraphics[width=\linewidth]{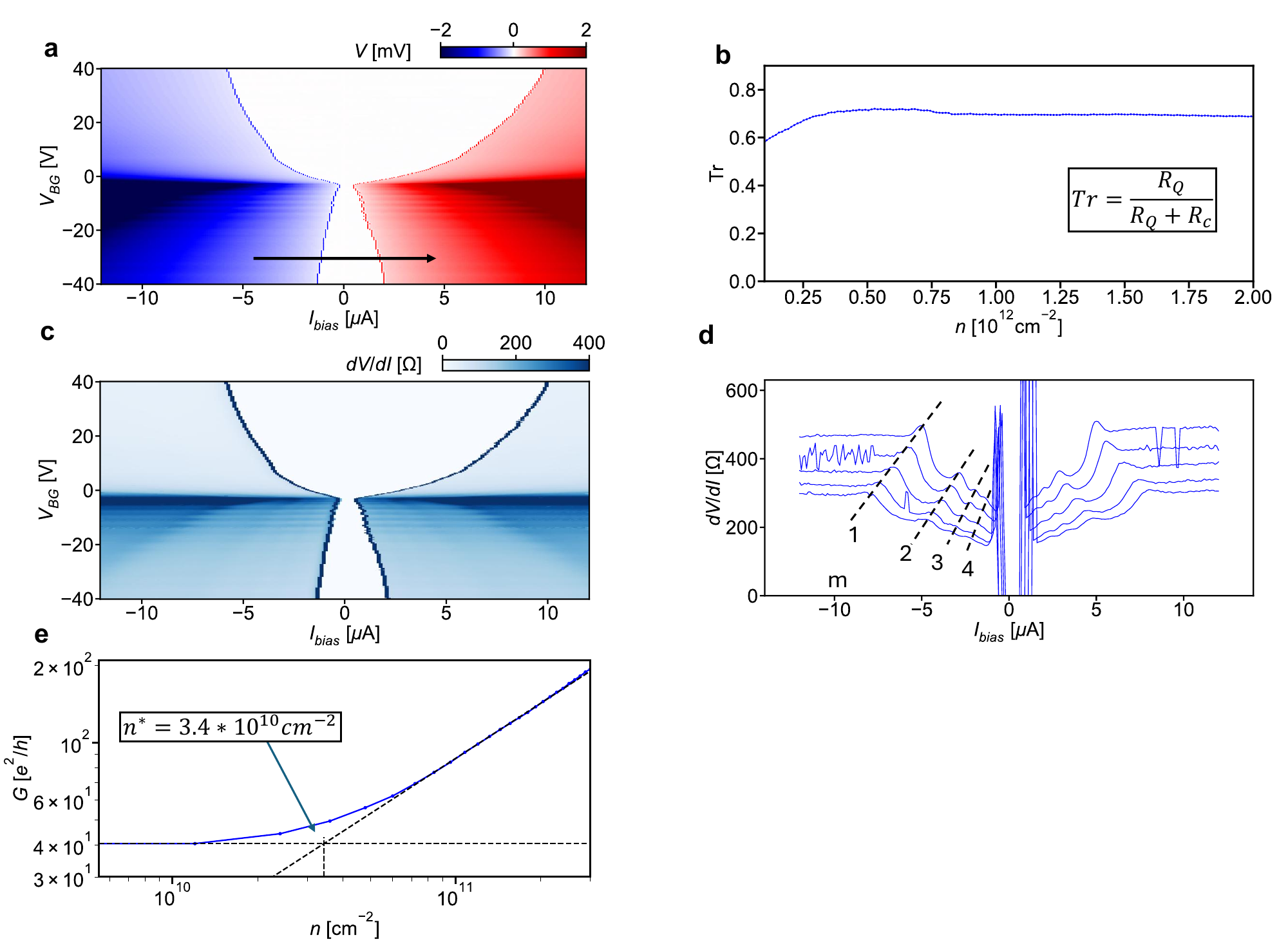}
	\caption{Additional characterization of device D1. ($\mathbf{a}$) Voltage drop $V$ as a function of DC current bias $I_{bias}$ and back-gate voltage $V_{BG}$ at $B = 0$ T, $T=40$~mK. The current sweep direction is from negative to positive bias (left to right), as indicated by the black arrow. The dotted blue and red lines correspond respectively to the retrapping and switching currents. ($\mathbf{b}$) Interface transparency $Tr$ as a function of carrier density $n$, calculated from the normal state resistance. $R_Q$ is the quantum ballistic limit, $R_c$ the contact resistance. ($\mathbf{c}$) Differential resistance $dV/dI$ as a function of $V_{BG}$ and $I_{bias}$, obtained from panel \textbf{a} by numerical differentiation. ($\mathbf{d}$) Selected $dV/dI$ curves around the charge neutrality point, with visible peaks due to MAR for $m=1-4$. ($\mathbf{e}$) $Log_{10}(G)$-$Log_{10}(n)$ plot for estimation of residual carrier density $n^*=3.4\times 10^{10}$cm$^{-2}$. \label{SI_fig_S1_charact_data}}
\end{figure*}

Here we report additional experimental data relative to device D1. Fig.~\ref{SI_fig_S1_charact_data}a shows a colormap of the voltage drop $V$ across the junction as a function of DC current bias $I_{bias}$ and back-gate voltage $V_{BG}$. The white region indicates zero voltage drop and identifies the supercurrent regime. The sweep direction of the current bias is indicated by the arrow and goes from negative to positive bias values. The blue and red dots indicate respectively the retrapping and switching currents, which are different, as explained in the main text. As for the normal state resistance (see Fig.~2a in the main text), also the supercurrent is modulated by the back-gate, reaching its maximum value at large positive $V_{BG}$ (in the electron doping regime) and its minimum at the charge neutrality point. To the left of the charge neutrality point ($V_{BG}<V_{BG}^{CNP}$), in the hole-doping regime, the supercurrent is always smaller compared to the electron doping side. This is due to the additional p-n interfaces that originate from n-type doping induced by Nb contacts and the resulting reduced interface transparency. In this regime, Fabry-Pérot oscillations are visible in the supercurrent, with local maxima in the switching current corresponding to local minima in the resistance. 

As explained in Ref.~\cite{BenShalom2016}, the interface transparency $Tr$ can be calculated from the normal-state resistance $R$ as the ratio $Tr=R_Q/(R_Q+R_c)$, where $R_Q=(h/e^2)/4N$ is the quantum ballistic limit, $N=\text{int}(2W/\lambda_F)$  is the number of propagating modes, $\lambda_F$ is the Fermi wavelength, the factor 4 accounts for the graphene degeneracy and $R_c = (R-R_Q)/2$ is the contact resistance. This method has been widely used to evaluate the interface transparency (see for example Refs.~\cite{Calado_2015,Borzenets2016}). It also provides a more reliable measurement as compared to the excess current, as the latter requires a large bias which heats up the sample (as discussed in Ref.~\cite{Amet2016}). As shown in Fig.~\ref{SI_fig_S1_charact_data}b, the interface transparency is not constant, but changes slightly with carrier concentration. A maximum $Tr=0.73$ is obtained for device D1.

An important benchmark for interface transparency in Josephson junctions is provided by the observation of Multiple Andreev Reflections (MARs). MAR features appear in the differential resistance at bias voltage values $V=2\Delta/me$, where $\Delta$ is the superconducting gap and $m$ an integer number. This can be seen in Fig.~\ref{SI_fig_S1_charact_data}c where MAR appear as darker stripes on the hole doping side. In Fig.~\ref{SI_fig_S1_charact_data}d we report selected $dV/dI$ curves as a function of $I_{bias}$ around the CNP, where MARs are indicated by the black dashed lines for $m=1$ to $m=4$. As discussed in Ref.~\cite{Kjaergaard2017}, for the n-th order Andreev reflection, the current amplitude $I(V)$ is proportional to $Tr^{m+1}$, so it is rapidly suppressed as the transparency $Tr$ decreases. In addition, we observe peaks in the differential resistance, as opposed to dips (see Fig.~\ref{SI_fig_S1_charact_data}d). This is typical for highly transparent junctions \cite{Kjaergaard2017}.

To further characterize device D1, we provide calculations of the residual carrier density (charge density fluctuations). Following procedures routinely used for graphene devices (see Ref.~\cite{Pezzini_2020}), we obtain the residual carrier density $n^*$ from the low temperature conductance $G$ as a function of the carrier density is $n=C_g (V_{BG}-V_{BG}^{CNP}$), where $C_g=6\times 10^{10}$cm$^{-2}$V$^{-1}$ is the back-gate lever arm. Specifically, from a $Log_{10}(G)$--$Log_{10}(n)$ plot, it is possible to extract $n^*$ as shown in Fig.~\ref{SI_fig_S1_charact_data}e. We obtain $n^*=3.4\times 10^{10}$~cm$^{-2}$.

\begin{figure*}[!htb]
	\centering
	\includegraphics[width=\linewidth]{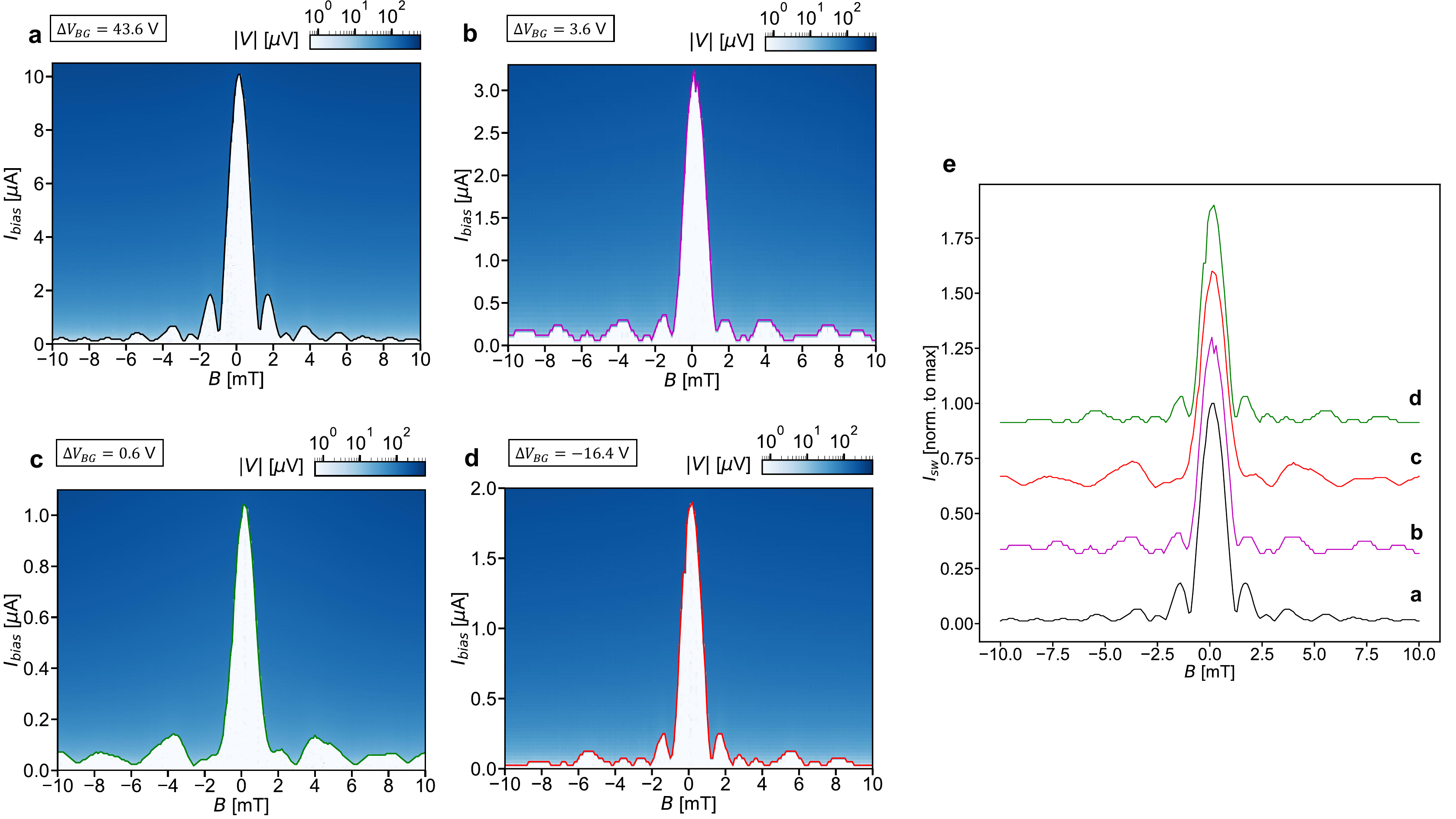}
	\caption{Extended Fraunhofer patterns from device D1. The bias sweep direction is from $0$ to positive bias. $T = 40$ mK for all acquisitions. Distance from CNP is indicated. ($\mathbf{a}$) $V_{BG}=40$~V. ($\mathbf{b}$) $V_{BG}=0$ V. ($\mathbf{c}$) $V_{BG}=-3$ V. ($\mathbf{d}$) $V_{BG}=-20$ V. ($\mathbf{e}$) Staggered plot of switching currents in \textbf{a-d} to highlight variations in the Fraunhofer shape.\label{SI_fig_S1_add_charact}}
\end{figure*}

Extended Fraunhofer patterns in the $\pm10$ mT range are reported in Fig.~\ref{SI_fig_S1_add_charact}a-d, corresponding to different $V_{BG}$ values and doping regimes ($V_{BG}=40$~V in \textbf{a}, $V_{BG}=0$~V in \textbf{b}, $V_{BG}=-3$~V in \textbf{c}, $V_{BG}=-20$~V in \textbf{d}). In this case, the current bias sweep direction is from zero to positive bias and the colored lines indicate the (positive) switching currents. In Fig.~\ref{SI_fig_S1_add_charact}e we compare the switching currents extracted from \textbf{a}-\textbf{d} and observe that Fraunhofer patterns have different shapes depending on the applied $V_{BG}$.

\subsection{Diode effect in the retrapping current}
As mentioned in the main text, switching and retrapping currents share the same rectification mechanism. Indeed, the lower value of the retrapping current when sweeping the DC current bias value from finite values to zero (starting from a finite resistance condition), is caused by a self-heating effect, as discussed in Ref.~\cite{Borzenets2016}.

\begin{figure*}[t]
	\centering
	\includegraphics[width=0.8\linewidth]{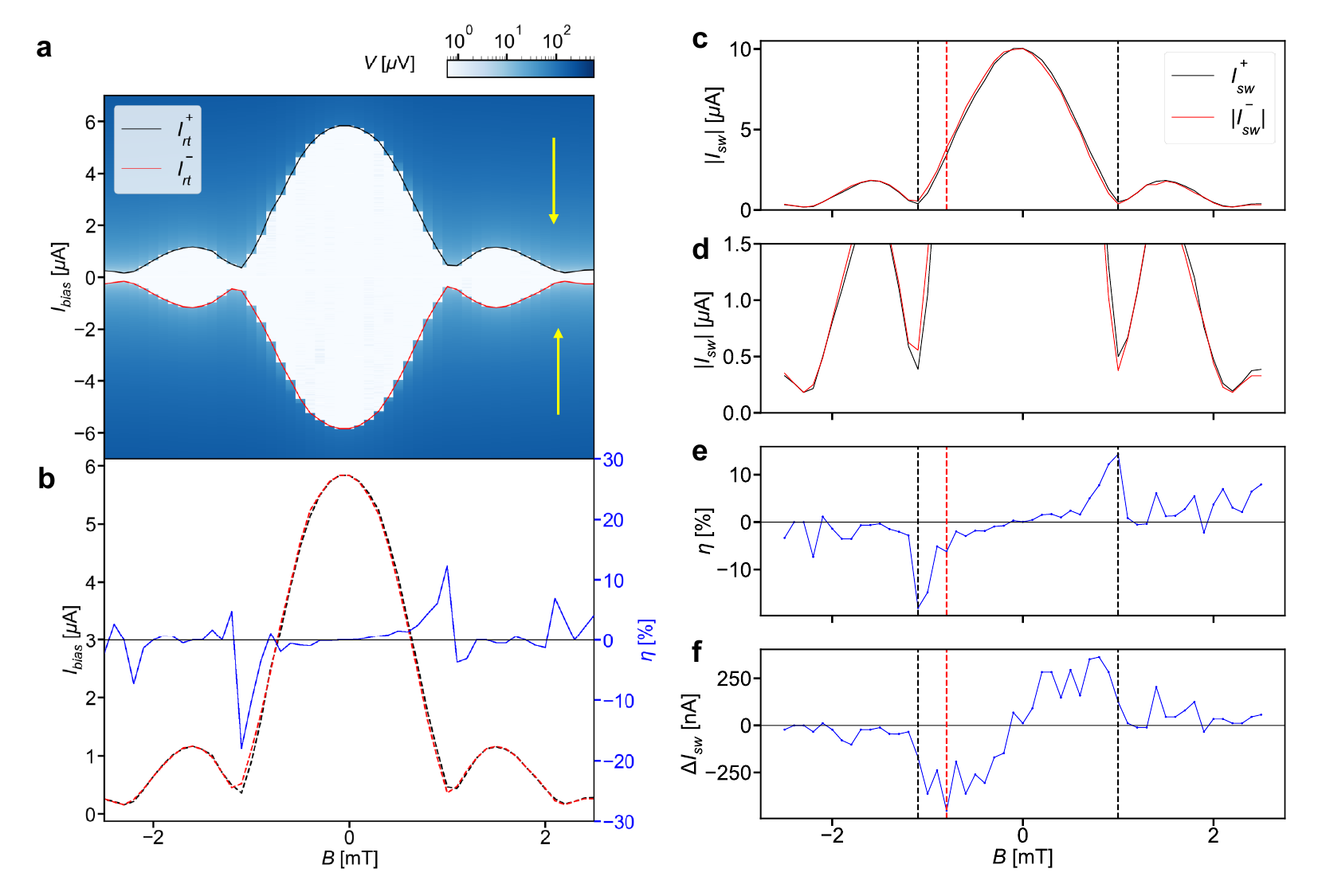}
	\caption{($\mathbf{a}$) Fraunhofer pattern obtained by plotting the retrapping branch of the current bias sweep, as indicated by the yellow arrows. $\Delta V_{BG}=43.6$~V, $n=2.6\times10^{12}$~cm$^{-2}$. ($\mathbf{b}$) Extracted values of positive and negative retrapping currents $I_{rt}^+$, $I_{rt}^-$, and calculated rectification coefficient $\eta$. ($\mathbf{c}$) Positive and negative switching currents, as extracted in Fig.~1c of the main text.  $\Delta V_{BG}=43.6$~V. ($\mathbf{d}$) Magnification of positive and negative switching currents to highlight differences at the Fraunhofer nodes, data from panel \textbf{a}. ($\mathbf{e}$) Rectification coefficient $\eta$, calculated as explained in main text. Maximum $\eta$ is observed in correspondence to the Fraunhofer nodes, as indicated by the vertical black dashed lines. ($\mathbf{f}$) Absolute value of the switching current difference $\Delta I_{sw}=I_{sw}^+-|I_{sw}^-|$. The maximum switching current difference $\Delta I_{sw}=-453$~nA is observed at $B=-0.8$~mT (vertical red dashed line), the corresponding $\eta=6.2\%$.  \label{SI_fig_D1_add_metrics}}
\end{figure*}

In Fig.~\ref{SI_fig_D1_add_metrics}a we report the Fraunhofer pattern at $\Delta V_{BG}=43.6$V resulting from retrapping current branches, i.e. measured by sweeping the DC current bias $I_{bias}$ from finite values to 0, as opposed to the switching current (yellow arrows indicate the bias sweep direction). A finite rectification coefficient is observed also in this case. The rectification coefficient, shown in Fig.~\ref{SI_fig_D1_add_metrics}b maintains the same characteristics as that calculated from the switching current: it is enhanced by the Fraunhofer nodes, it exhibits a characteristic antisymmetric shape around B=0 and the maximum rectification efficiency is similar and close to $20\%$.

\subsection{Additional metrics for the diode effect}

In the following we provide additional discussion on the diode effect on device D1. We focus on the Fraunhofer pattern at large electron doping shown in Fig.~1c of the main text. The corresponding switching currents are reported in Fig.~\ref{SI_fig_D1_add_metrics}c. The rectification coefficient $\eta$ is obtained from Fig.~\ref{SI_fig_D1_add_metrics}c and is shown in Fig.~\ref{SI_fig_D1_add_metrics}e.

In addition to the rectification coefficient, an important parameter to evaluate the performance of a diode is provided by the absolute value of the switching current difference $\Delta I_{sw}=I_{sw}^+-|I_{sw}^-|$, reported in Fig.~\ref{SI_fig_D1_add_metrics}f. Indeed, the rectification coefficient $\eta$ can in principle become large in correspondence to the Fraunhofer nodes due to the denominator $I_{sw}^++|I_{sw}^-|$ becoming small. However, in correspondence to the Fraunhofer nodes at $\Phi=\pm\Phi_0$, the critical current is not entirely suppressed (i.e. the Fraunhofer nodes do not close entirely). This is shown in the zoom-in presented in Fig.~\ref{SI_fig_D1_add_metrics}d: at the minima, both positive and negative switching currents are still of the order of several hundreds of nA. The rectification coefficient, that is clearly enhanced at the Fraunhofer nodes (vertical black dashed lines Fig.~\ref{SI_fig_D1_add_metrics}e), exhibits its maximum $\eta=18.0\%$ at $\Phi=-\Phi_0$. The corresponding switching current values are $I_{sw}^+=385$~nA, $I_{sw}^-=-555$~nA, resulting in $\Delta I_{sw}=I_{sw}^+-|I_{sw}^-|=-170$~nA. Comparable values are observed for the Fraunhofer patterns presented in Fig.~2 of the main text. The metric provided by the rectification coefficient $\eta$ remains therefore valid up to $I_{sw}\sim 350$~nA.

For the Fraunhofer pattern presented in Fig.~1c, the maximum $\Delta I_{sw}=-453$~nA is observed for a magnetic flux $\Phi=-0.75\Phi_0$  ($B=-0.8$~mT, vertical red dashed line in Fig.~\ref{SI_fig_D1_add_metrics}f). While maximum $\eta$ is observed in correspondence to the Fraunhofer nodes, the maximum switching current difference $\Delta I_{sw}$ is observed within the main lobe and closer to $B=0$. At this magnetic field, the device supports positive and negative switching currents $|I_{sw}|>3\;\mu$A. The corresponding rectification efficiency $\eta=6.2\%$ is lower than the maximum value at the nodes but still sizeable.

In Table \ref{tab:mod_coeff} we summarize the maximum rectification efficiency and the corresponding values of $I_{sw}^+$, $|I_{sw}^-|$, $\Delta I_{sw}$, $B$ for increasing minimum current threshold values, for the Fraunhofer pattern in Fig.~\ref{SI_fig_D1_add_metrics}a. Analogous considerations are valid also for all Fraunhofer patterns presented in Fig.~2 of the main text.

\begin{table}[!htb]
	\begin{center}
		\begin{tabular}{c|c|c|c|c|c}
			$I_{sw}$ threshold [$\mu$A] & Max $\eta$ & $I_{sw}^+$ [$\mu$A] & $|I_{sw}^-|$ [$\mu$A] & $\Delta I_{sw}$ [nA] & Corresponding $B$ [mT] \\
			\hline
			0.1 & $18.0\%$ & $0.385$ & $0.555$ & $-170$ & $-1.1$ \\
			0.3 & $18.0\%$ & $0.385$ & $0.555$ & $-170$ & $-1.1$ \\
			0.5 & $14.8\%$ & $1.043$ & $1.405$ & $-363$ & $-1.0$ \\
			2.5 & $6.2\%$ & $3.423$ & $3.876$ & $-453$ & $-0.8$ \\
			5.0 & $2.4\%$ & $6.267$ & $5.973$ & $295$ & $0.6$ \\
		\end{tabular}
		\caption{\label{tab:mod_coeff} Maximum rectification efficiency, corresponding $I_{sw}^+$, $|I_{sw}^-|$, $\Delta I_{sw}$ and magnetic field $B$, for different thresholds on the switching current.}
	\end{center}
\end{table}

\subsection{Diode effect at the charge neutrality point}

Among the patterns shown in Fig.~\ref{SI_fig_S1_add_charact}, the one in c stands out as it shows stronger deviations from the ideal $sin(x)/x$ dependence, compared to the others. Specifically, the central lobe merges with the first lateral ones, and higher order lobes are irregular in shape and width. This pattern was taken at small distance from the charge neutrality point, {$\Delta V_{BG}=0.6$~V}, $n\sim 10^{10}\;cm^{-2}$, in a transport regime which is dominated by charge puddles.

\begin{figure*}[t]
	\centering
	\includegraphics[width=\linewidth]{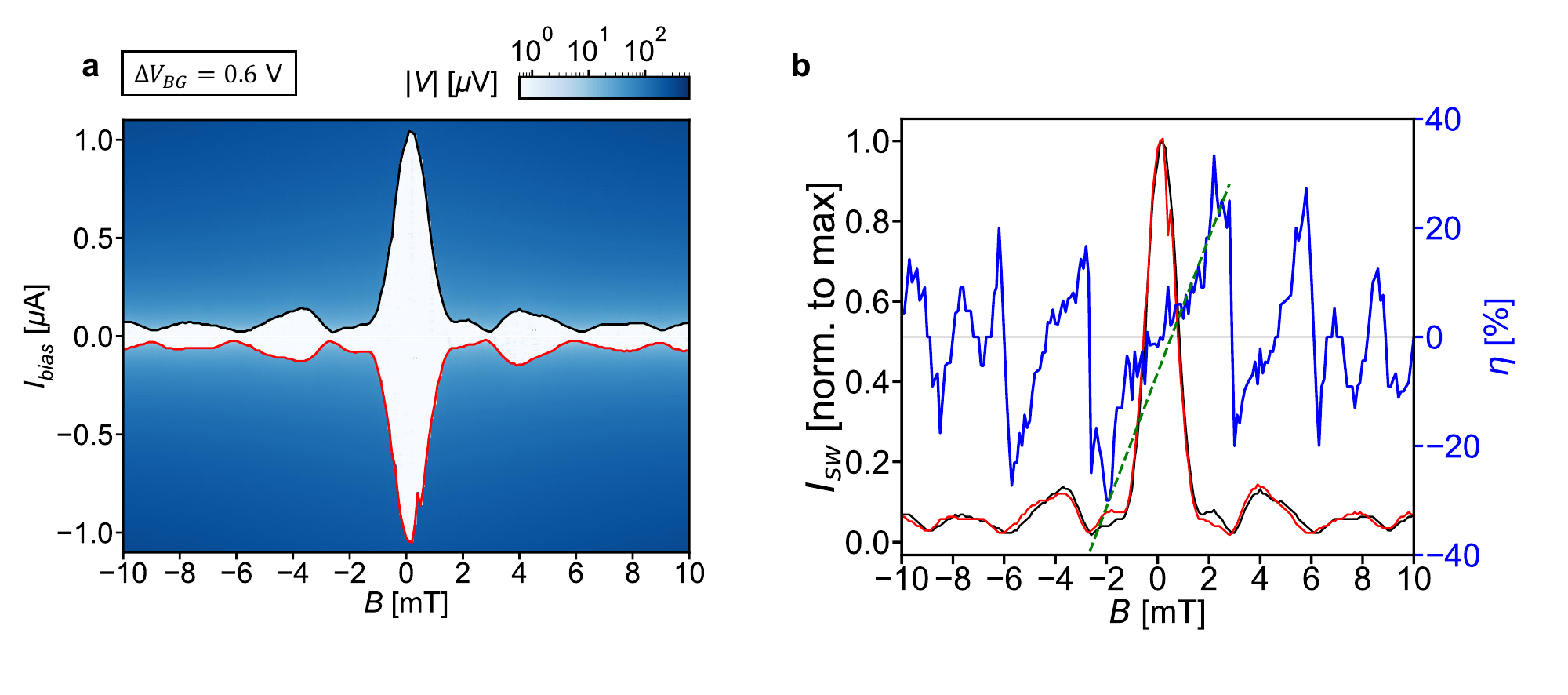}
	\caption{($\mathbf{a}$) Fraunhofer pattern for device D1, $V_{BG}=-3$ V, $T = 40$ mK.   ($\mathbf{b}$) Rectification coefficient for data in \textbf{a}. \label{SI_fig_S1_JDE_CNP}}
\end{figure*}

In Fig.~\ref{SI_fig_S1_JDE_CNP}a, the pattern at $V_{BG}=-3$~V ($\Delta V_{BG}=0.6$~V) is shown for both positive and negative current direction, along with the rectification coefficient $\eta$ in Fig.~\ref{SI_fig_S1_JDE_CNP}b. The rectification coefficient exceeds $20\%$ and is antisymmetric around $B=0$. The strong deviations from the ideal shape of the Fraunhofer pattern can be qualitatively linked to variations in the supercurrent distribution. However, there is no reason to expect a suppression of the diode effect, since it is originated by an asymmetry in the electrostatic potential, which is present even in low doping regimes. Both in the experiment and in the simulations, the rectification coefficient is not suppressed close to the CNP, but $\eta$ exhibits shape, amplitude and magnetic field dependence comparable to the highly doped case.

\clearpage
\section{Device D2}
In the following we report data relative to device D2 that was fabricated some months after sample D1 using the same fabrication protocol. Fabrication steps are described in the Methods section of the main text.

\begin{figure*}[!htb]
	\centering
	\includegraphics[width=0.9\linewidth]{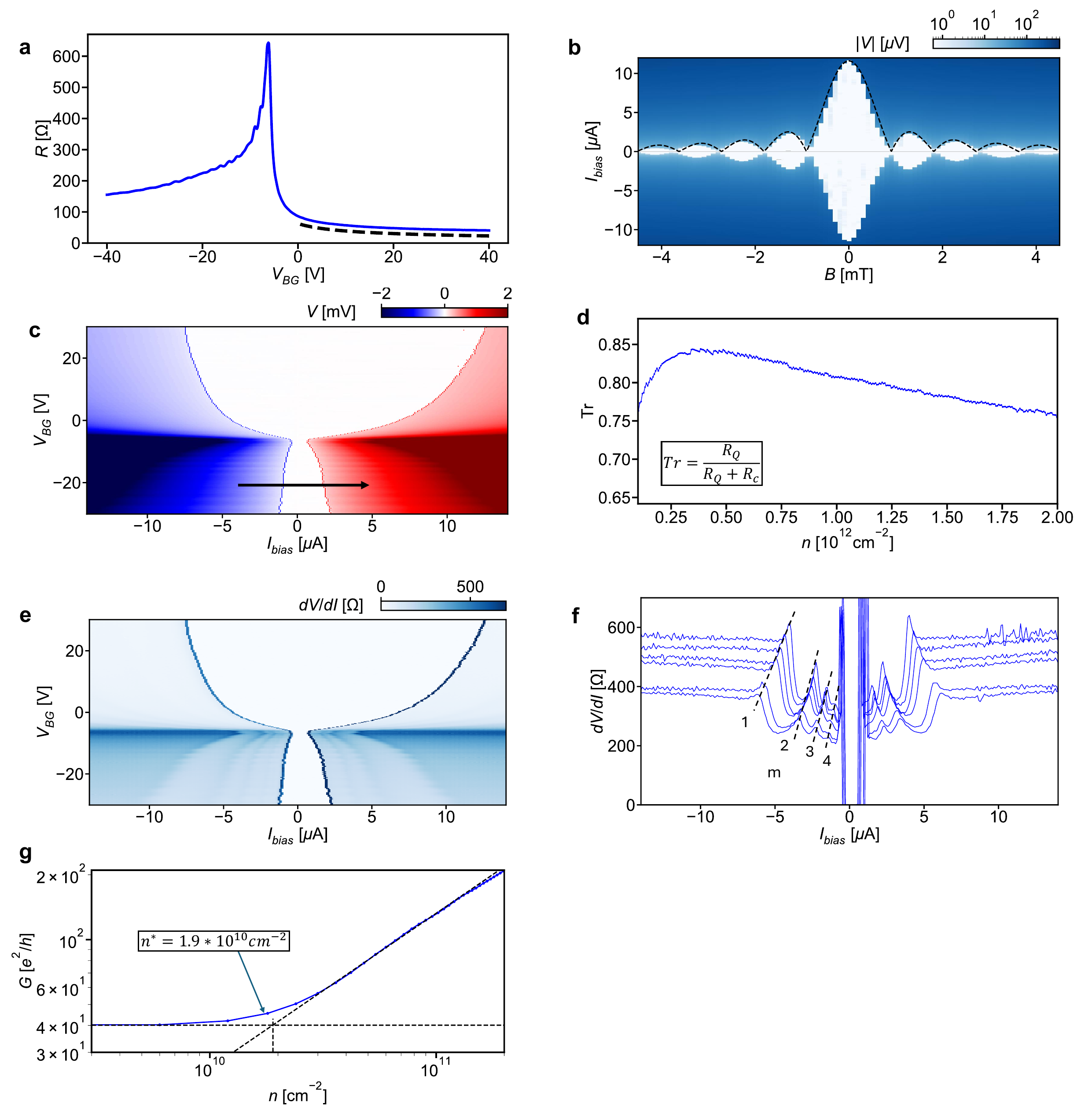}
	\caption{Characterization of device D2. ($\mathbf{a}$) Resistance $R$ as a function of back-gate voltage $V_{BG}$. In the electron doping regime, the resistance approaches the Sharvin limit (black dashed line). $V_{CNP}=-6.2$ V in this device. $T=2.7$~K. ($\mathbf{b}$) Extended Fraunhofer pattern, $V_{BG}=30$ V ($\Delta V_{BG}=36.2$ V). The bias sweep direction is from $0$ to either positive or negative bias, showing therefore positive and negative switching currents. The black dashed line is the ideal $sin(x)/x$ pattern. ($\mathbf{c}$) Voltage drop V as a function of DC current bias $I_{bias}$ and back-gate voltage $V_{BG}$ at $B = 0$ T. The current sweep direction is from negative to positive bias (left to right), as indicated by the black arrow. The dotted blue and red lines correspond respectively to the retrapping and switching currents. $T = 350$ mK for acquisitions in b and c. ($\mathbf{d}$) Interface transparency $Tr$ as a function of carrier density $n$, calculated from the normal state resistance. ($\mathbf{e}$) Differential resistance $dV/dI$ as a function of back-gate voltage $V_{BG}$ and DC current bias $I_{bias}$. MAR peaks on the hole-doping side appear as dark stripes. ($\mathbf{f}$) Selected $dV/dI$ curves around the charge neutrality point, with visible peaks due to MAR for $m=1-4$. ($\mathbf{g}$) $Log_{10}(G)$-$Log_{10}(n)$ plot for estimation of residual carrier density, $n^*=1.9\times 10^{10}$~cm$^{-2}$.\label{SI_fig_S2_charact}}
\end{figure*}

\subsection{Characterization}

Compared to device D1, device D2 exhibits consistent properties in both normal and superconducting regimes, as illustrated in Fig.~\ref{SI_fig_S2_charact}. The normal state resistance as a function of the back-gate voltage $V_{BG}$, shown in Fig.~\ref{SI_fig_S2_charact}a, approaches the Sharvin limit in the n-type doping regime, indicating large interface transparency. An "extended" Fraunhofer pattern is reported in Fig.~\ref{SI_fig_S2_charact}b. Compared to the Fraunhofer pattern of device D1 in the same doping regime (Fig.~1c of main text), it shows minor deviations from the ideal pattern (black dashed line).
The supercurrent is modulated by the BG voltage, as shown in Fig.~\ref{SI_fig_S2_charact}c, and reaches its maximum value for large electron doping, with a supercurrent density close to $4\;\mu$A/$\mu$m. Fig.~\ref{SI_fig_S2_charact}d shows the interface transparency $Tr$ as a function of charge carrier density $n$, calculated as explained in Section S1.1. As compared to device D1, device D2 exhibits a larger interface transparency, $\max(Tr)=0.84$. In the differential resistance map of Fig.~\ref{SI_fig_S2_charact}e, obtained from numerical differentiation of Fig.~\ref{SI_fig_S2_charact}c, MAR features can be clearly seen as darker stripes in the differential resistance on the hole-doping side. Selected $dV/dI$ showing MAR peaks are reported in Fig.~\ref{SI_fig_S2_charact}f: MAR peaks are more pronounced compared to device D1, consistent with a larger interface transparency \cite{Kjaergaard2017}. Device D2 also exhibits lower charge density fluctuation amplitude, $n^*=1.9\times 10^{10}$~cm$^{-2}$ as shown in Fig.~\ref{SI_fig_S2_charact}g.

\subsection{Diode effect in device D2}
\begin{figure*}[!htb]
	\centering
	\includegraphics[width=\linewidth]{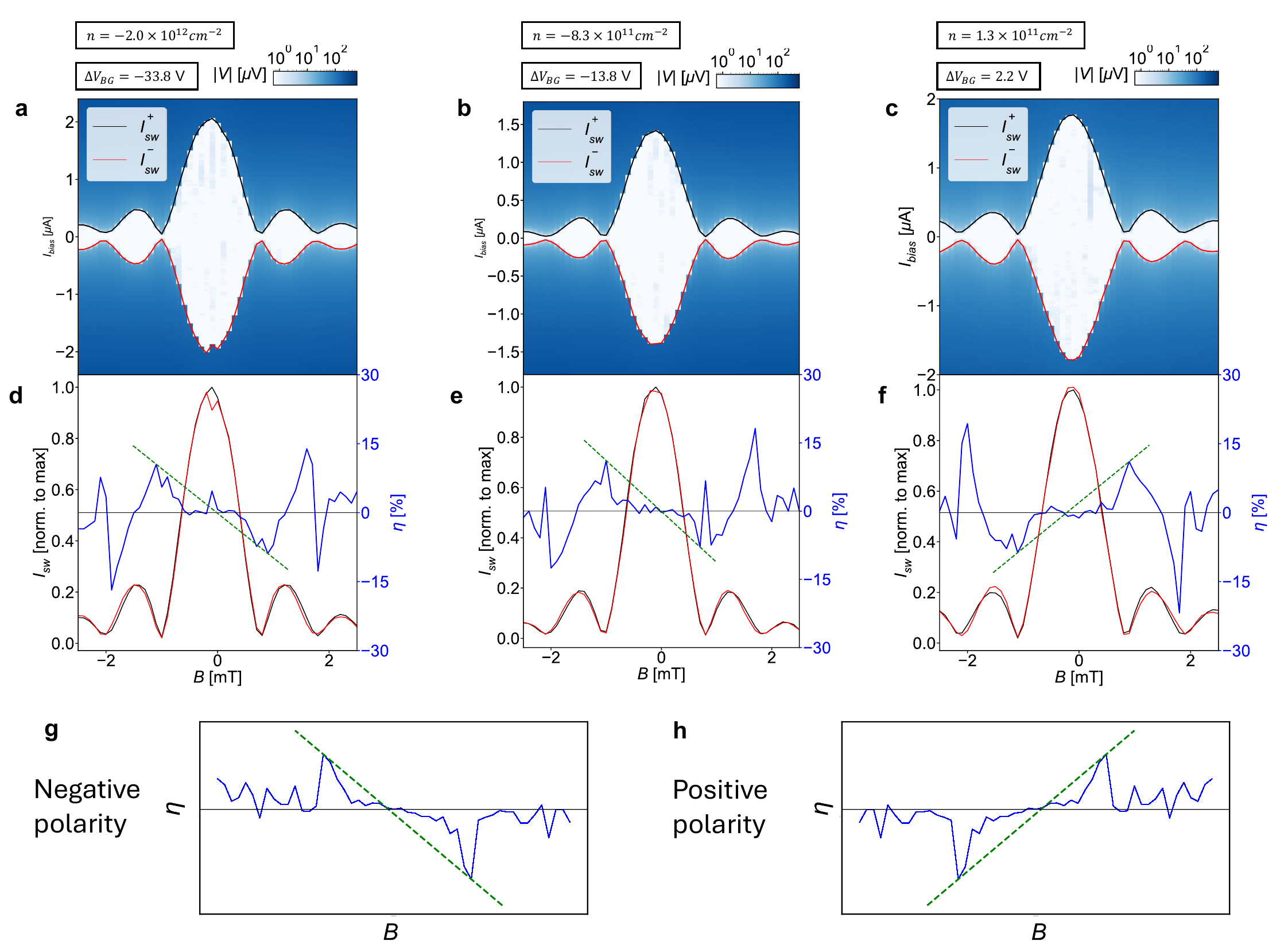}
	\caption{Fraunhofer patterns of device D2 (voltage drop $|V|$ as a function of out-of-plane magnetic field $B$ and DC bias current $I_{bias}$) and corresponding rectification parameter $\eta$ at different back-gate voltage values. The relative position to the charge neutrality point (CNP) $\Delta V_{BG}=V_{BG}-V_{CNP}$ and the corresponding charge carrier density values are indicated ($V_{CNP}=-6.2$~V). (\textbf{a}-\textbf{d}) $V_{BG}=-40$~V. (\textbf{b}-\textbf{e}) $V_{BG}=-20$~V. (\textbf{c}-\textbf{f}) $V_{BG}=-4$~V. (\textbf{g}-\textbf{h}) Sketch of the definition of diode polarity. We define the diode polarity as the sign of the slope of the line obtained by connecting peaks in the rectification coefficient $\eta$. Regarding the experimental data, the diode in \textbf{f} has opposite (positive) polarity compared to \textbf{d},\textbf{e}. \label{fig_SI_FH_S2_diode}}
\end{figure*}

Among all tested BG voltage values, ranging from $+40$ V to $-40$ V and comprising both n-type and p-type doping regimes, device D2 exhibits supercurrent rectification only at some BG voltage values. These Fraunhofer patterns are shown in Fig.~\ref{fig_SI_FH_S2_diode}a-c, and correspond respectively to $V_{BG}=-40$ V, $V_{BG}=-20$ V, $V_{BG}=-4$ V. The relative distance to the CNP is indicated at the top of each Fraunhofer pattern. As for device D1 presented in the main text, supercurrent rectification is observed in the main lobe, with enhancement at first nodes of the Fraunhofer pattern. In all three cases $\eta \neq 0$ also on first lateral lobes.

However, device D2 shows different diode polarities as the BG voltage is changed, differently from device D1 which always exhibits the same diode polarity regardless of the BG voltage. The slope of the green dashed line (traced by connecting peaks in $\eta$ at first nodes) in Fig.~\ref{fig_SI_FH_S2_diode}f is opposite compared to Figs.~\ref{fig_SI_FH_S2_diode}d,e. 

As discussed in the main text, a non-$\mathbf{M_x}$-symmetric scattering potential is necessary for the diode effect to appear, and the details of the scattering potential affect some of the features observed in the rectification coefficient. Changes in the diode polarity are expected as the scattering potential changes, though this happens in an unpredictable way and specifically without direct control by the back-gate voltage. 
The Fraunhofer patterns of device D2 are characterized by a much more regular shape and are closer to the ideal shape (black dashed line in Fig.~\ref{SI_fig_S2_charact}b). It is reasonable to assume that, in the case of a rather uniform potential landscape,  small asymmetries become relevant only when the doping level is low enough, or due to the additional p-n interfaces (originating from the interplay of induced doping and Fermi level pinning at Nb interfaces). Supercurrent rectification in device D2 is indeed observed in the hole doping regime (Fig.~\ref{fig_SI_FH_S2_diode}a,b) and for low electron doping (Fig.~\ref{fig_SI_FH_S2_diode}c), while it is absent for large electron doping.

\clearpage
\section{Diode effect in device D3 with different geometry}

\begin{figure*}[!htb]
	\centering
	\includegraphics[width=0.8\linewidth]{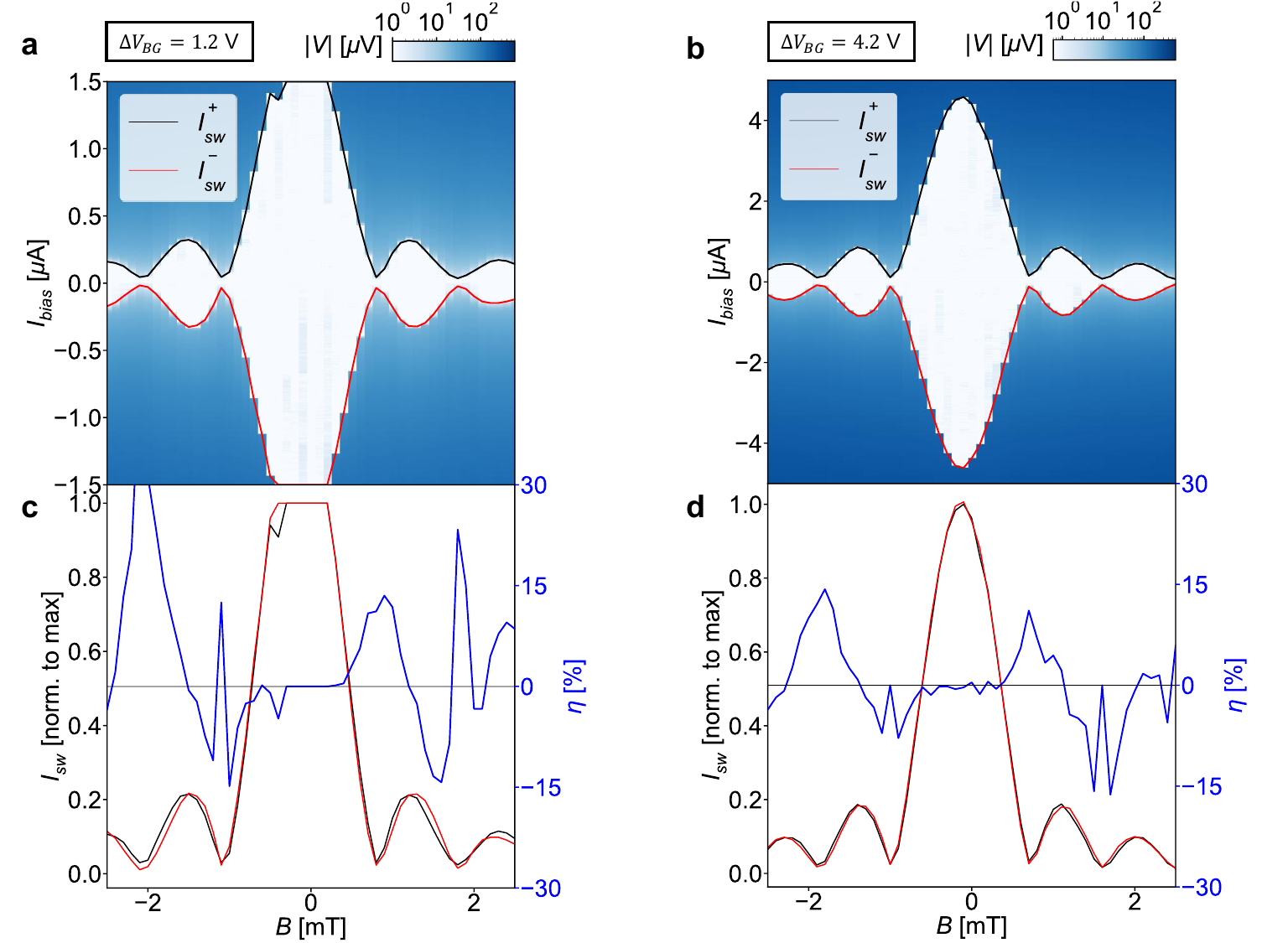}
	\caption{Fraunhofer patterns of device D3 (voltage drop $|V|$ as function of out-of-plane magnetic field $B$ and DC bias current $I_{bias}$) and corresponding rectification parameter $\eta$ at different back-gate voltage values. $T=350$~mK. (\textbf{a}-\textbf{c}) $V_{BG}=-3$~V. (\textbf{b}-\textbf{d}) $V_{BG}=0$~V. \label{fig_SI_FH_D3}}
\end{figure*}

Measurements from an additional device D3 with different geometry ($L=600$~nm, $W=3\;\mu$m) are presented in Fig.~\ref{fig_SI_FH_D3} for two different back-gate voltage values. Despite the different geometry of D3, the supercurrent rectification exhibits similar properties as for devices D1 and D2: the rectification coefficient presents a characteristic antisymmetric shape around B=0, it is enhanced at Fraunhofer nodes, and a finite rectification is also reported on lateral lobes.

\clearpage
\section{Scattering matrix formulation}

The theoretical description of the diode effect in the Fraunhofer pattern of planar Josephson junctions has been provided in Ref.~\cite{chirolli2025} and arises due to the breaking of mirror symmetry of the junction due to the scattering potential in the junction. Here we report a summary of the scattering matrix model used in this work. For the case of graphene, we assume that a pattern of electron and hole puddles is present in the graphene sheet, which induces a potential $V(x,y)$ that electrons experience while traversing the junction. For the particular setup, we assume superconducting leads attached to the top and the bottom of the scattering region, thus defining the $y$ direction, and that the potential is not symmetric under the mirror symmetry about the plane orthogonal to the $x$ direction, such that $V(-x,y)\neq V(x,y)$. 

The equilibrium currents through the junction can be expressed by means of the free energy of the system  $F(\varphi)$ at temperature $T$, $I(\varphi)=\frac{2\pi}{\Phi_0}\frac{\partial F(\varphi)}{\partial \varphi}$ \cite{bardeen1969},
where $\varphi$ is the phase difference across the junction. Depending on the direction of the applied bias current, we can define two critical currents as
\begin{eqnarray}
	I_{c}^+=\max_{\varphi} I(\varphi),\qquad
	I_{c}^-=\min_{\varphi} I(\varphi),
\end{eqnarray}
that can be different, yielding a diode effect, that is quantified by the rectification coefficient
\begin{equation}
	\eta=\frac{I_{c}^+-|I_{c}^-|}{I_{c}^++|I_{c}^-|}.
\end{equation}
At zero temperature, the free energy is obtained in terms of the Andreev spectrum, which can be obtained by means of the  unitary scattering matrix $s(\epsilon)$ describing scattering through the junction at energy $\epsilon$, by solving the eigenproblem \cite{beenakker1991,beenakker1997,beenakker2004}
\begin{equation}
	s_{\rm A}s_{\rm N}\psi=\psi,
\end{equation}
where the normal (N) and Andreev (A) scattering matrices are given by
\begin{equation}
	s_{\rm N}=\left(\begin{array}{cc}
		s(\epsilon) & 0\\
		0 & s^*(-\epsilon)
	\end{array}\right),
	\qquad 
	s_{\rm A}=\alpha \left(\begin{array}{cc}
		0 & r_A^*\\
		r_A & 0
	\end{array}\right),
\end{equation}
where  $r_A={\rm diag}
(e^{i\varphi/2},e^{-i\varphi/2})$, and  $\alpha=e^{-i~{\rm arccos}(\epsilon/\Delta)}$, with $\Delta$ the superconducting gap of the terminals. In the Andreev approximation of a scattering region of linear size much smaller than the coherence length $\xi=\hbar v_F/\Delta$, we can neglect the energy dependence of the scattering matrix $s(\epsilon)$, so that $s(\epsilon)=s(-\epsilon)\equiv s$. Following Ref.~\cite{vanheck2014}, the Andreev spectrum is particle-hole symmetric by construction, and the positive eigenvalues are determined by the non-negative singular values of the matrix 
\begin{equation}
	A=\frac{1}{2}(r_As+s^T r_A).
\end{equation}
In presence of time-reversal symmetry, the scattering matrix is symmetric, $s^T=s$. Furthermore, if mirror symmetry about the plane orthogonal to the plane of the junction and containing the phase bias direction is preserved, $M_x^{-1}sM_x$, it can be shown that no diode effect arises \cite{chirolli2025}. It turns out that the high transparency of the junction and a potential that breaks the mirror symmetry defined by $M_x$ are sufficient to yield a diode effect at finite magnetic field \cite{chirolli2025}.

\clearpage
\section{Diode polarity flip}

\begin{figure*}[!htb]
	\centering
	\includegraphics[width=\linewidth]{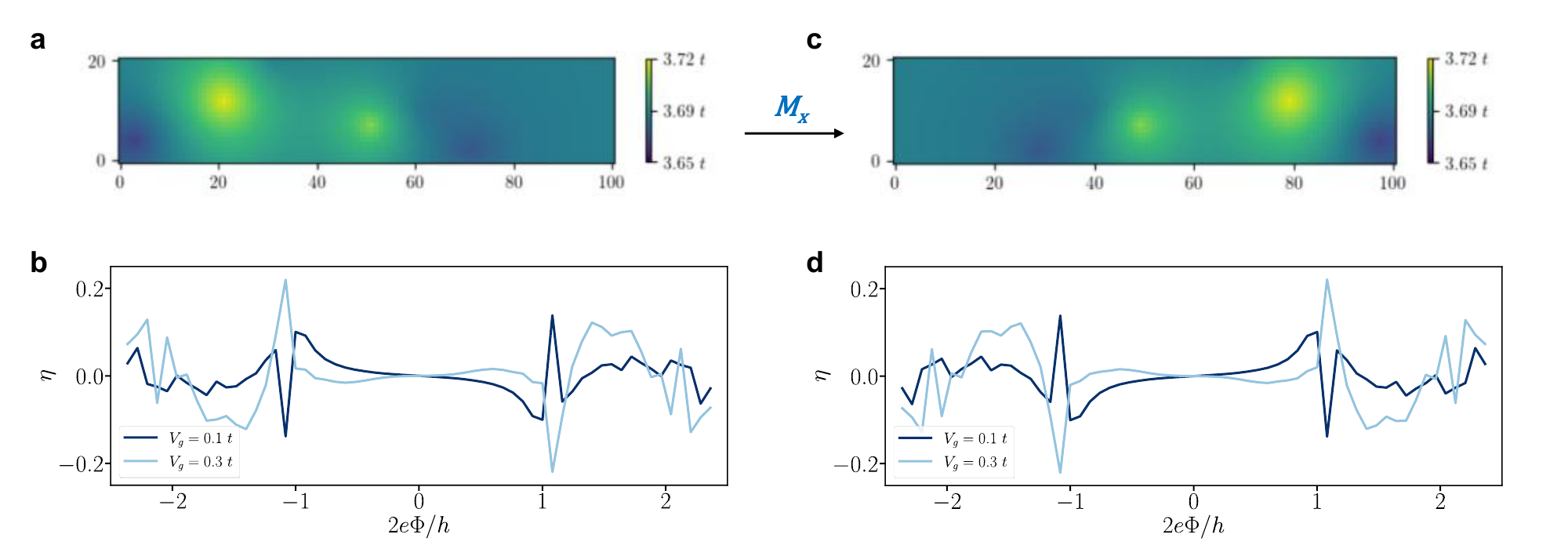}
	\caption{Effect of mirror symmetry of the scattering potential on the diode rectification coefficient. (\textbf{a}) Onsite potential. (\textbf{b}) Simulated rectification coefficient $\eta$ for the scattering potential shown in \textbf{a}, for two gate voltage values. (\textbf{c}) Onsite potential obtained by performing a mirror symmetry operation about the $\bf{M_x}$-axis on the potential shown in \textbf{a}. (\textbf{d}) Simulated rectification coefficient $\eta$ for the scattering potential shown in \textbf{c}, for the same gate voltage values as in \textbf{b}. The rectification coefficient flips sign. 
		\label{fig_mirror_flip}}
\end{figure*}

Here we discuss the possibility to control the diode polarity by acting on the scattering potential. As illustrated in Fig.~\ref{fig_mirror_flip}, a diode polarity flip is obtained by the theoretical model when a mirror symmetry operation about the $\bf{M_x}$-axis (see Fig.~1a of the main text) is performed on the scattering potential. In Fig.~\ref{fig_mirror_flip}b we report the rectification coefficient for the scattering potential in (a) for two example values of gate voltage, while results shown in Fig.~\ref{fig_mirror_flip}d correspond to the scattering potential in (c). The latter was obtained from the potential in (a) by performing a mirror symmetry operation about the $\bf{M_x}$-axis. The rectification coefficient shown in (d) is the same as in (b), but with opposite sign, indicating a polarity flip. 

As previously discussed, no changes in the diode polarity are reported experimentally for device D1, while we observe a polarity flip for device D2. Since there is no direct control on the scattering potential by acting only on the back-gate, no deterministic control of the diode polarity is obtained in our devices. However, such control could in principle be implemented by using additional side gates, which have been demonstrated to be highly effective in tuning transport properties in Josephson junctions both in the normal and dissipation-less regimes \cite{Seredinski2019, Turini2025}.

\clearpage
\bibliographystyle{unsrt}
\bibliography{Bibliography_paper_SDE} 

\end{document}